\documentclass[sigconf]{acmart}

\usepackage{amsmath}
\usepackage{algorithm}
\usepackage[noend]{algpseudocode}
\usepackage[T1]{fontenc}
\usepackage[utf8]{inputenc}
\usepackage{multirow}
\usepackage[toc,page]{appendix}
\usepackage{adjustbox}
\usepackage{makecell}
\usepackage{hyphenat,fancyref,hyperref,xspace,tabularx,enumitem}
\usepackage[labelformat=simple,skip=0pt]{subcaption}
\usepackage{calc,booktabs,xcolor,url,hyperref}
\usepackage[normalem]{ulem}

\makeatletter
\def\BState{\State\hskip-\ALG@thistlm}
\makeatother

\newcommand{\ie}{{\itshape i.e.}\xspace}
\newcommand{\eg}{\emph{e.g.}\xspace}
\newcommand{\etc}{\emph{etc.}\xspace}

\DeclareMathOperator*{\argmax}{arg\,max}

\fancyrefchangeprefix{\fancyreffiglabelprefix}{f}
\fancyrefchangeprefix{\fancyreftablabelprefix}{t}
\fancyrefchangeprefix{\fancyrefeqlabelprefix}{eqn}

\graphicspath{{../}{Figures/}}

\hypersetup{colorlinks=true, urlcolor=blue,
    citecolor=black, linkcolor=black, menucolor=black,
    draft=false}

\newcommand{\parabreak}{\vspace*{1.00ex minus 0.25ex}\noindent}
\renewcommand{\paragraph}[1]{\vspace*{1ex plus 0.25ex minus 0.25ex}\noindent {\bfseries #1}}

\newcommand{\systemname}{{\sf In-N-Out}\xspace}
\newcommand{\systemnameposs}{{\sf In-N-Out's}\xspace}

\newcommand{\textred}[1]{\textcolor{red}{#1}}
\ifx\noeditingmarks\undefined
   \newcommand{\pgwrapper}[2]{\textred{#1: #2}}
\else
   \newcommand{\pgwrapper}[2]{}
\fi

\AtBeginDocument{%
  \providecommand\BibTeX{{%
    \normalfont B\kern-0.5em{\scshape i\kern-0.25em b}\kern-0.8em\TeX}}}

\copyrightyear{2020}
\acmYear{2020}
\setcopyright{acmcopyright}
\acmConference[MobiCom '20]{The 26th Annual International Conference on Mobile Computing and Networking}{September 21--25, 2020}{London, United Kingdom}
\acmBooktitle{The 26th Annual International Conference on Mobile Computing and Networking (MobiCom '20), September 21--25, 2020, London, United Kingdom}
\acmPrice{15.00}
\acmDOI{10.1145/3372224.3380899}
\acmISBN{978-1-4503-7085-1/20/09}

\begin{document}

\title{Towards Flexible Wireless Charging for Medical Implants Using Distributed Antenna System}


\author{
Xiaoran Fan$^{\ast, \dagger}$, Longfei Shangguan$^{\ddagger}$, Richard Howard$^{\dagger}$, Yanyong Zhang$^{\ast}$
}
\author{
Yao Peng$^{\star}$, Jie Xiong$^{\diamond}$, Yunfei Ma$^{\times}$, Xiang-Yang Li$^{\ast}$
}
\affiliation{\normalsize
   \institution{$^{\ast}$University of Science and Technology of China, $^{\dagger}$Rutgers University, $^{\ddagger}$Microsoft\\
   $^{\star}$Northwest University,
   $^{\diamond}$UMASS Amherst,
   $^{\times}$Alibaba Group}}

\renewcommand{\shortauthors}{X. Fan, L. Shangguan, R. Howard, Y. Zhang, Y. Peng, J. Xiong, Y. Ma, and X. Li}

\begin{abstract}
This paper presents the design, implementation and evaluation of \systemname, a software-hardware solution for far-field wireless power transfer.
\systemname can continuously charge a medical implant residing in deep tissues at near-optimal beamforming power, even when the implant moves around inside the human body.
To accomplish this, we exploit the unique energy ball pattern of distributed antenna array and devise a backscatter-assisted beamforming algorithm that can  concentrate RF energy on a tiny spot surrounding the medical implant. Meanwhile, the power levels on other body parts stay in low level, reducing the risk of overheating.
We prototype \systemname on 21 software-defined radios and a printed circuit board (PCB).
Extensive experiments demonstrate that \systemname achieves 0.37~mW average charging power inside a 10~cm-thick pork belly, which is sufficient to wirelessly power a range of commercial medical devices.
Our head-to-head comparison with the state-of-the-art approach shows that \systemname achieves 5.4$\times$--18.1$\times$ power gain when the implant is stationary, and 5.3$\times$--7.4$\times$ power gain when the implant is in motion.
\end{abstract}



\maketitle

\section{Introduction}
\label{s:introduction}

Each year millions of patients improve their quality of life through medical implants~\cite{khan2014implantable}.
These devices are inserted into the human body to
replace a missing body part~\cite{Prosthesis}, modify a body function~\cite{Cochlear}, 
or provide supports to organs and tissues~\cite{Transvaginal}.
While functional innovations on medical implants are going full steam ahead, the amount of energy required by these devices remains substantial.
Though cutting-edge batteries could enable medical implants (\eg, pacemaker~\cite{Pacemaker}) to function for years~\cite{nuclearBattery,rasouli2010energy}, the use of battery is not always feasible -- there may not be enough space inside the brain or body as a battery's size is proportional to its lifetime~\cite{armand2008building}.
We have thus seen cumbrous solutions such as placing the battery of a brain stimulator in the user's chest or even outside the body, with wires running between the battery and the stimulator.
Battery replacement, on the other hand, is risky as it usually requires a surgery that may introduce extra complications~\cite{pacemakerBatteryProblem,gul2011common}.

\begin{figure}[t]
\centering
{\includegraphics[width=0.9\columnwidth]{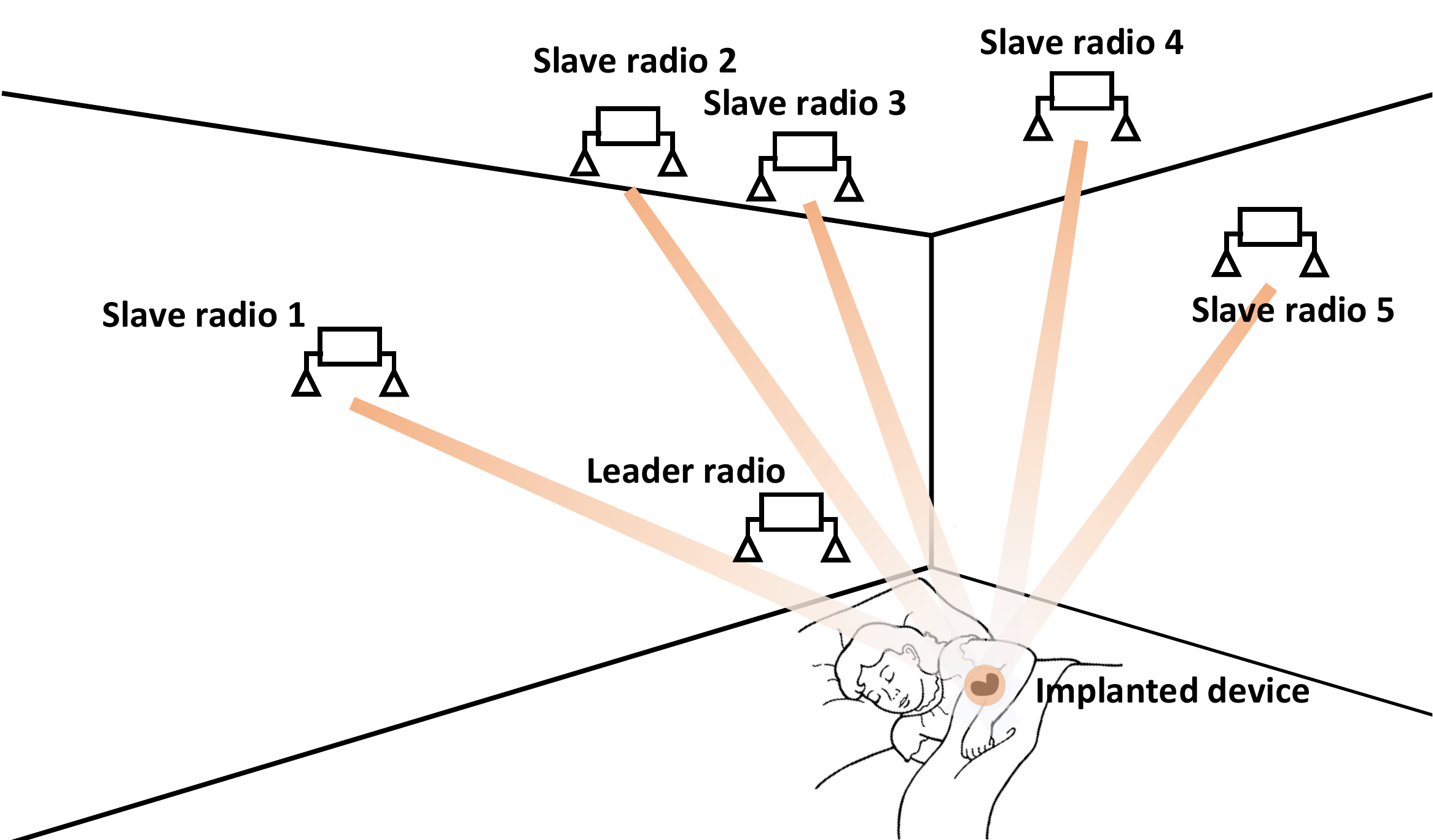}}
\caption{\textbf{An illustration of \systemname deployment}. The leader radio coordinates multiple slave radios to charge the pacemaker during bedtime.}
\label{fig:system}
\end{figure}

Wireless charging has received attention in recent years as a viable alternative.
The concept of wireless charging, however, is not new.
From early 1900s Tesla's Wardenclyffe tower~\cite{Wardenclyffe} to the later Air Force mission of powering an unmanned helicopter~\cite{brown1965experimental}, wireless charging has witnessed significant breakthroughs over the past century.
Today wireless charging can be simply performed on an office desk or in a car.
As far as medical implants are concerned, they are primarily charged through electromagnetic coupling in the near field~\cite{schuler1961high,lenaerts2007inductive,jow2007design,donelan2008biomechanical,ramrakhyani2011design}.
These near-field charging systems use dedicated coils that usually require contact with human tissues.
A critical drawback of these systems is that their charging efficiency drops significantly with the reduction of coil size and the increase of coil separation, which severely hinders the miniaturization of medical implants~\cite{ho2014wireless}.
Another drawback of these near-field systems is the low flexibility: the users are required to wear bulky transmitter coils and carefully align them with the implant coils~\cite{standStill}.
Even though the user can stay static for hours, the inter-coil coupling can be easily broken as the implant coils may move as blood flows~\cite{ImplantsandProsthetics}.
Thus, a contactless means of wireless charging holds appeal as a flexible and less invasive alternative.

\parabreak This paper presents \systemname, a flexible far field power transfer system that owns two desirable properties: 1) it does not require the user to wear cumbersome charging devices. 2) it can continuously charge the medical implant residing in deep tissues with consistently near-optimal power, even when the implant moves around inside the human body.
To do so, 
\systemname leverages \emph{beamforming} to combine signals coherently at the medical implant.
At the heart of beamforming is the accurate measurement of channel state information (CSI) of each wireless channel. This is usually achieved by having the transmitter send a preamble, where the receiver (\eg, a medical implant) uses this preamble to estimate the CSI of the forward channel.
This CSI value is then fed back for transmitter beamforming.

However, CSI measurement becomes very challenging, if at all possible, for medical implants. RF signal generation is power hungry, which becomes especially challenging for medical implants that are deeply power constrained~\cite{ho2014wireless}.
In practice, to minimize power consumption, the RF radio of a medical implant typically adopts a rather low power amplification coefficient~\cite{hannan2014energy}.
Therefore, the resulting preamble signals are very weak, whi\-ch are made even worse by the fast decaying  radiation efficiency of an in-body antenna. The antenna's radiation efficiency decays significantly due to its miniature size, i.e., 10 -- 20~dB loss compared to the weak transmission signals~\cite{kim2004implanted,surowiec1987vitro}.
Furthermore, RF signals experience exponentially more attenuation in human tissues than in air,
\eg, 40~dB loss over just a few centimeters in muscles~\cite{dove2014analysis}.
As a result, the received signal is usually well below the noise floor, hence the failure to provide accurate CSI estimation.

 \begin{figure}[t]
      \centering
      \includegraphics[width=\columnwidth]{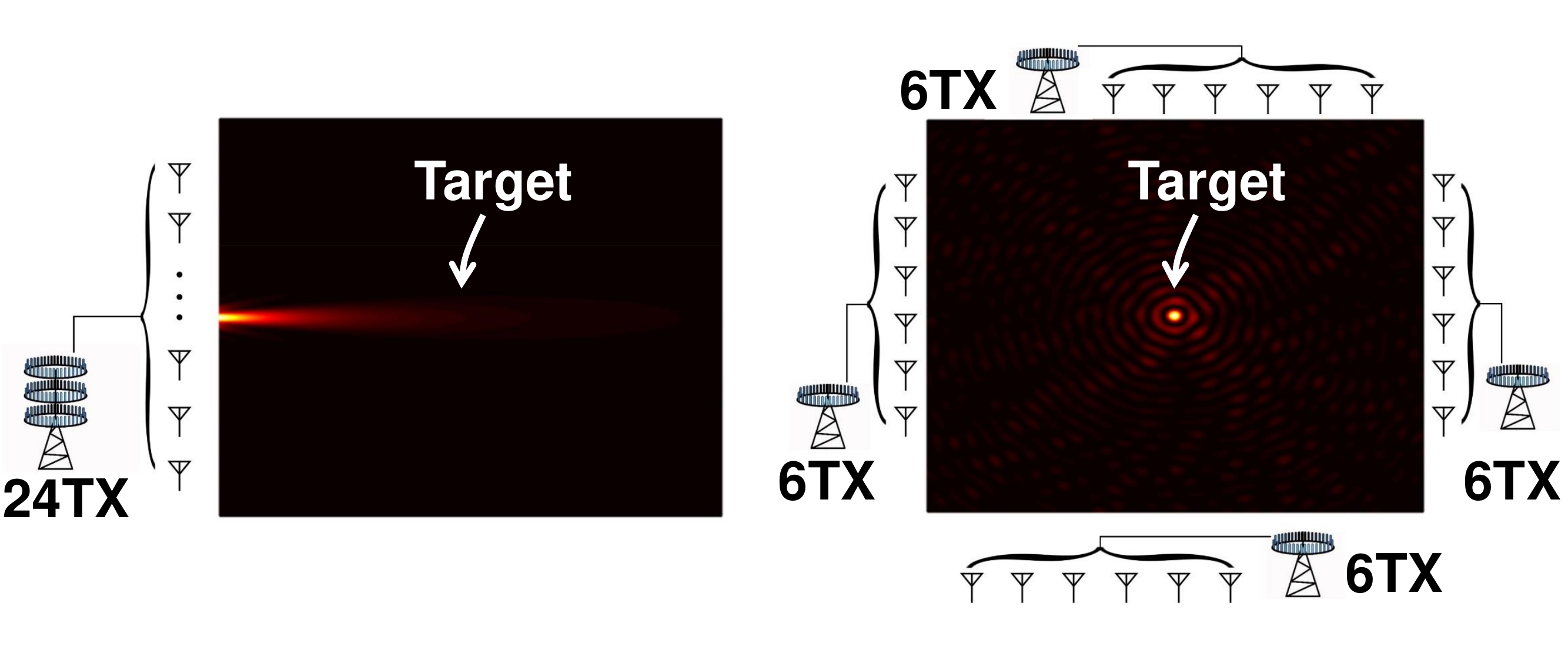}
      \caption{\textbf{The energy heatmap produced by (left) a linear 24-antenna array and (right) a distributed 24-antenna array.} 
  The linear antenna array produces an energy beam spreading in the direction of the target, while the distributed antenna array produces an energy spot surrounding the target.
 }
 \label{fig:simu_algo}
 \end{figure}

To solve these challenges, the state-of-the-art, IVN~\cite{yunfei}, proposes to encode the frequency of multiple transmission signals in hopes of these frequency-varying signals coherently combine at the medical implant from time to time, without CSI measurements.
IVN achieves high beamforming power intermittently to cold start the medical implant.
It is, however, ill\hyp{}suited for power transfer as the beamforming power it actually produces, for most of the time, is far less than the \emph{maximal} beamforming power.\footnote{Defined as the power level measured at the target location when all wireless transmissions are coherently combined.} The coherent-incoherent beamforming nature renders the power delivery particularly inefficient.

In \systemname, we devise a coherent beamforming algorithm that can continuously achieve the maximal beamforming power at the medical implant, even when the implant moves around inside the human body. Our algorithm builds upon the iterative one-bit phase alignment approach proposed in~\cite{mudumbai2005scalable,mudumbai2006distributed}, which involves the receiver sending a feedback signal to describe the received beamforming power change after each iteration until reaching the maximal. Though this approach can accomplish consistently coherent beamforming, it cannot be directly adopted in our setting because frequently measuring beamforming power and sending feedback signals would even consume more power than what can be wirelessly harvested at the implant. Thus, leveraging the one-bit phase alignment approach as a generic framework, we take into consideration the unique challenges in our scenario and design a backscatter assisted beamforming (in short, BAB) scheme. Our BAB scheme employs a customized monotonic \emph{backscatter} radio at the implant that simply reflects signals and another nearby auxiliary radio that assesses the received backscatter signal power change. In this way we successfully offload power-consuming operations at the medical implant (\eg, power assessment, signal generation and transmission) to the auxiliary radio outside human body, and thus significantly cut down the energy consumption compared to existing systems where implant radios have to directly assess the received power change and produce feedback signals. 

However, new challenges arise when we use backscatter radios at the implants. After going through excessive channel fading in both directions, the received backscatter signal is usually well below the noise floor, hence causing the new challenge of detecting/decoding the weak backscatter signal.
\systemname addresses this challenge by pre-coding the carrier signal using chirp spreading spectrum (CSS) modulation.
The frequency-domain processing gain of CSS enables \systemname to detect the backscatter signal even 35dB\footnote{\systemname does not need to decode the packet but detect the power change of backscatter signals.} below the noise floor.

We prototype \systemname on 21 USRP software defined radios and evaluate its performance in various settings. 
In our prototype, we adopt distributed antenna layout that addresses the safety concerns of wireless charging. Performing beamforming using co-located antennas will generate a high energy beam along a specific angle, as shown in Figure 2(left). This high energy beam does not only cover the medical implant but also part of the human body, likely resulting in excessive heating of human tissues. In contrast, beamforming with distributed antennas produces a tiny energy spot surrounding the target location as shown in Figure 2(right) and the energy density at other locations is significantly lower due to destructive interference~\cite{fan2018secret,fan2018energy, fan2018enabling, fan2019facilitating}. Therefore, it naturally avoids overheating other areas of the body while charging the target. Moreover, these distributed antennas have different orientations and are thus insensitive to the orientation of the implant.

Our field studies show that \systemnameposs beamforming algorithm is efficient (< 0.3~s latency) and reliable (insensitive to the implant's rotation and motion).
It achieves 0.37~mW charging power on average when the implant is 2~m away, which is sufficient to power a range of medical devices from outside the body.
Our head-to-head comparison with IVN~\cite{yunfei} shows that \systemname achieves 5.4$\times$--18.1$\times$ and 5.3$\times$--7.4$\times$ average power gain over IVN in stationary and low-speed mobile scenarios, respectively.

\systemnameposs contributions include:

\begin{itemize}
    \item Designing a software-hardware solution for deep tissue power transfer. We devise a set of signal processing algorithms and a low-power, monotonic backscatter radio that enables \systemname to charge the medical implant at the maximal beamforming power, even when the implant moves around inside the human body. Our system consists of several technical innovations, including 
    backscatter-leader-slave three-party beamforming without explicit CSI measurement, two-phase le\-ader-slave chirp synchronization design, radio cold start through intentionally imperfect phase alignment, \etc 
    
    
    \item Prototyping the system on software\hyp{}defined radios and a PCB board, and conducting comprehensive evaluation of the system. Our evaluation takes into consideration the impact of important parameters such as the charging medi\-um, system size, chirp bandwidth, antenna array size, \etc We also conduct head-to-head comparisons with the state-of-the-art approach in a range of settings.  
\end{itemize}

In the next section (\S\ref{s:uc}), we introduce the scope of our work. We introduce beamforming primer in Section~\ref{s:pre}.
The system design is detailed in Section~\ref{s:design}. An
implementation (\S\ref{s:implementation}) and performance evaluation (\S\ref{s:evaluation}) then follow. 
Section~\ref{s:related_works} summarizes related works. 
We discuss future works in Section~\ref{s:disc} and conclude the work in
Section~\ref{s:concl}. 

\vspace{-0.1cm}
\section{Scope}
\label{s:uc}

This work aims to developing a practical wireless charging system, with the hope of extending the lifetime of medical implants.

The lifetime of a medical implant depends mainly on the lifetime of its battery~\cite{lifetimeMedical}.
Hence a lot of efforts have been made to improve the battery life~\cite{mallela2004trends}, either by increasing the battery capacity or minimizing the device power consumption.
Today state-of-the-art pacemakers can last for over ten years~\cite{batterypace}. However, the user still needs a surgery for replacement when the battery is depleted.
To lengthen the implant's lifetime,  \systemname can serve as a supplementary power supply -- whenever the user stays in a space where a personalized \systemname system is available, the implant can be charged, without drawing power from the regular battery. As a result, the lifetime of the implant can be significantly extended.

We note that though \systemname is primarily designed for wireless power transfer, its application scope can be much broader.
For example, \systemname could potentially serve as a communication system to collect the biomedical data from inbody sensors~\cite{cha2012cmos,kiourti2012review}.
Compared with conventional gastroscopy that requires the patient to swallow a tube for data collection~\cite{Gastroscopy}, our solution is much less invasive.


\paragraph{Possible Deployment Scenarios}. We envision the \systemname system will be deployed in the user's personal space (home and/or office).
Given a typical bedroom (4$\times$4~$m^2$ rectangular area with a 2.8~m average target-antenna distance), if we keep the number of radios to a reasonable number, i.e., less than 14 (each emitting 30~dBm signals), then the resulting power density at any location in the room is well below the power limit specified by FCC regulation (0.6 $mw/cm^2$~\cite{fcc}).  

\section{Beamforming Without CSI Feedback}
\label{s:pre}
\label{ss:design_beamformingWithoutCSI}

Due to excessive channel fading and inhomogeneous channel propagation in deep tissues, CSI measurement becomes very challenging, if at all possible, for medical implants embedded in deep tissues (will be explained in \S\ref{sss:carrierSignal}).
Instead of pursuing a precise CSI measurement, we employ a non-CSI beamforming approach proposed in~\cite{mudumbai2005scalable,mudumbai2006distributed},  referred as one-bit phase alignment algorithm. 

\paragraph{Algorithm overview}. The one-bit phase alignment algorithm goes through multiple rounds and then converges to the optimal phase settings.
In each round, each transmitter updates the phase of the transmission signal based on the feedback sent from the receiver.
The phase value in the current round is randomly selected within the range of $\pm \Phi$ with respect to the phase value in the previous round (we discuss the optimal $\Phi$ setting in \S\ref{sss:optPhaseAdjust}.).
Phase update can be formulated as follows:
\begin{equation*}\small
	\begin{aligned}
      &\theta_{i}(n+1)=
      \begin{cases}
      \theta_{i}(n)+ \delta_{i}(n), & \text{if}\ y[n]> y[n-1], -\Phi^{\circ} \leq \delta_{i}(n) \leq \Phi,\\
      \theta_{i}(n-1) + \delta_{i}(n), & \text{otherwise},
    \end{cases}
	\end{aligned}
\end{equation*}
where $\theta_{i}(n)$ is the phase setting of the $i^{th}$ transmitter in the $n^{th}$ round and
$y[n]$ is the received signal power in the $n^{th}$ round.

\begin{figure}[t]
\centering
{\includegraphics[width=1\columnwidth]{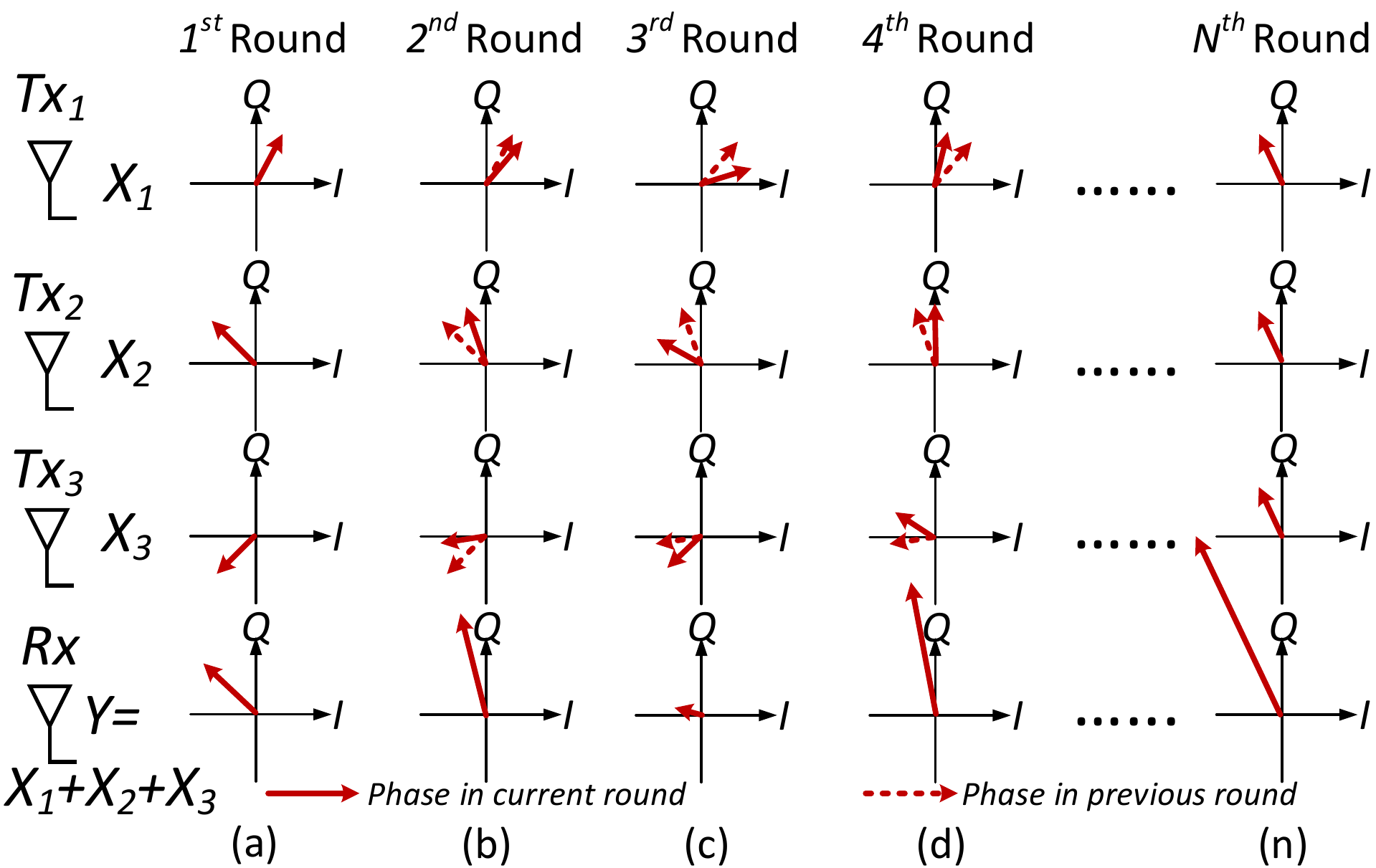}}
\caption{\textbf{A running example of the one-bit phase alignment algorithm with three transmitters}. Each transmitter adjusts its phase based on the feedback from the receiver and gradually converges to the optimal phase alignment.}
\label{fig:nocsibf}
\end{figure}

\paragraph{An example}. We use a simple example scenario involving three transmitters to explain this algorithm.
In the first round, each transmitter randomly chooses a phase value, as shown in Figure~\ref{fig:nocsibf}(a).
The receiver records beamforming power.
In the second round, each transmitter randomly chooses a phase that is within the range of $\pm \Phi$ from its phase value in the first round. This new phase leads to an increased beamforming power, as shown in Figure~\ref{fig:nocsibf}(b).
The receiver notifies transmitters this power increase ($\uparrow$) with a single bit feedback.
Hence in the third round, each transmitter uses its round two phase value as the reference and updates its phase accordingly, which unfortunately leads to a degraded beamforming power ($\downarrow$), shown in Figure~\ref{fig:nocsibf}(c).
Therefore, in the fourth round, each transmitter again uses its round two phase value as the reference (Figure~\ref{fig:nocsibf}(d)).
The algorithm iterates in the fashion until the beamforming power reaches its maximum (Figure~\ref{fig:nocsibf}(n)). 


\section{System Design}
\label{s:design}

\systemname involves a \emph{leader} radio and several \emph{slave} radios working on 915~MHz ISM band,\footnote{Working on 2.4~Ghz or 5~Ghz ISM band may introduce severe interference to ongoing Wi-Fi traffic, whereas working on lower ISM band~(\ie 433~MHz) requires a bulky receiving antenna which is not suited for implant devices.} as shown in Figure~\ref{fig:system}.
The leader node detects and decodes feedback signals sent from the medical implant and uses decoded information to govern the phase alignment of slaves in the next round.
As the medical implant may move around while charging, we do not assume any prior knowledge of the implant's location.
To minimize the energy consumption due to feedback signal creation and transmission, we design a low-power backscatter radio that offloads the computation from the medical implant to the leader radio that is outside of the human body (\S\ref{ss:design_bsAssistedBeamforming}).
As a proof of concept, we use a dedicated radio (\ie USRP) as the leader radio. However, we envision the leader radio can simply be a smartphone being able to talk with slave radios wirelessly (\eg, through Wi-Fi).

In the rest of this section, we explain the details of each design component, including carrier signal design and synchronization (\S\ref{sss:carrierSignal}), low-power backscatter radio design (\S\ref{sss:backscatterRadio}), and power change inference algorithm (\S\ref{sss:powerTrendInference}).
Finally, we explain the way to bootstrap the system during the cold start in \S\ref{ss:coldStart} and discuss the way to balance beamforming convergence and delay in~\S\ref{sss:optPhaseAdjust}.

\begin{table*}[t]
  \centering
  \begin{adjustbox}{width=2.1\columnwidth}
    \begin{tabular} {ccccccccc} \toprule[2pt]
      \makecell{\bf Tx power\\\bf (dBm)} & 
      \makecell{\bf Air path loss (dBm)\\\bf dist. (1 -- 10m)} & 
      \makecell{\bf Skin reflection\\\bf /absorption (dBm)} & 
      \makecell{\bf Muscle path loss (dBm)\\\bf dist. (2 -- 6cm)}&
      \makecell{\bf Insertion loss\\\bf (dBm)} &
      \makecell{\bf Muscle path loss (dBm)\\\bf dist. (2 -- 6cm)} &
      \makecell{\bf Skin reflection\\\bf /absorption (dBm)} &
      \makecell{\bf Air path loss (dBm)\\\bf dist. (1 -- 10m)} &
      \makecell{\bf Rx power\\\bf (dBm)}\\ 
      \midrule[1pt]
      30~\cite{fcc} & 
      31.67 -- 51.67 ~\cite{tse2005fundamentals}& 
      3\cite{rani2013transmission} & 
      9.2 -- 27.6~\cite{stango2016characterization,kim2011rf} & 30~\cite{zhang2016enabling} & 
      9.2 -- 27.6 ~\cite{stango2016characterization,kim2011rf} & 5\cite{rani2013transmission} & 
      31.67 -- 51.67 ~\cite{tse2005fundamentals}& 
      -89.74 -- -166.54\\
      \bottomrule[2pt]
    \end{tabular}
  \end{adjustbox}
  \caption{\textbf{The power loss at different part of the round-trip path between the transmitter (outside body) and the receiver (inside body)}.
  The transmission power is set to the maximum value under FCC regulation.}
  \label{tab:pathloss}
\end{table*}

\subsection{Backscatter Assisted Beamforming (BAB)}
\label{ss:design_bsAssistedBeamforming}

Directly applying one-bit phase alignment algorithm to in-body wireless charging is unfeasible due to its excessive energy overhead.
Generating a feedback signal with even the simplest modulation scheme (\ie frequency shift keying (F\-SK)) costs at least tens of milliwatts~\cite{hu2016braidio}, which can quickly add up when we go through each iteration.
This operation alone would consume more power than what can be wirelessly delivered to the implant.
To address this dilemma, we replace the default active radio on the medical implant with a low-power backscatter radio. 
A backscatter radio, while being able to minimize the implant's power consumption, raises new challenges and complicates the system design nonetheless.  
Below we discuss these challenges in detail as well as our solutions.

\subsubsection{Carrier Signal}
\label{sss:carrierSignal}

Backscatter radio neither generates carrier signals nor amplifies transmission signals. 
It instead modulates data on top of the ambient carrier signal (a sinusoidal tone coming from a nearby active radio) and reflects the modulated signal (termed as backscatter signal) back to the receiver.
Compared with the active radio, the backscatter radio saves three to four orders of magnitude transmission power by avoiding power consumption on carrier generator and power amplifier~\cite{talla2017lora}.
However, the lack of power amplifier renders the backscatter signal extremely weak, which is then made much worse by the excessive fading in deep tissues.
Table~\ref{tab:pathloss} shows the break-down signal attenuation as the carrier signal goes through the human body and reflected by the backscatter radio. 
The receiving power is around -128~dBm on average, well below the ambient RF noise floor measured by an USRP N210 in the same frequency band.\footnote{-70~dBm on 915~MHz frequency band measured in an office building.}
Hence, both CSI and RSS measurement are unreliable for channel estimation (CSI measurement at 915MHz band requires at least 6~dB higher signal strength than RSS measurement~\cite{hervas2015narrowband}).

An intuitive approach is to have the backscatter radio leverage more advanced coding mechanisms
to improve the signal to noise ratio (SNR) of the backscatter signal.
However, this requires more complicated, power hungry analog-to-digital (ADC) and digital circuits and will again complicate the implant radio design and boost the overall energy consumption.

\begin{figure}[t]
\centering
\begin{subfigure}[b]{0.495\linewidth}
\includegraphics[width=1\textwidth]{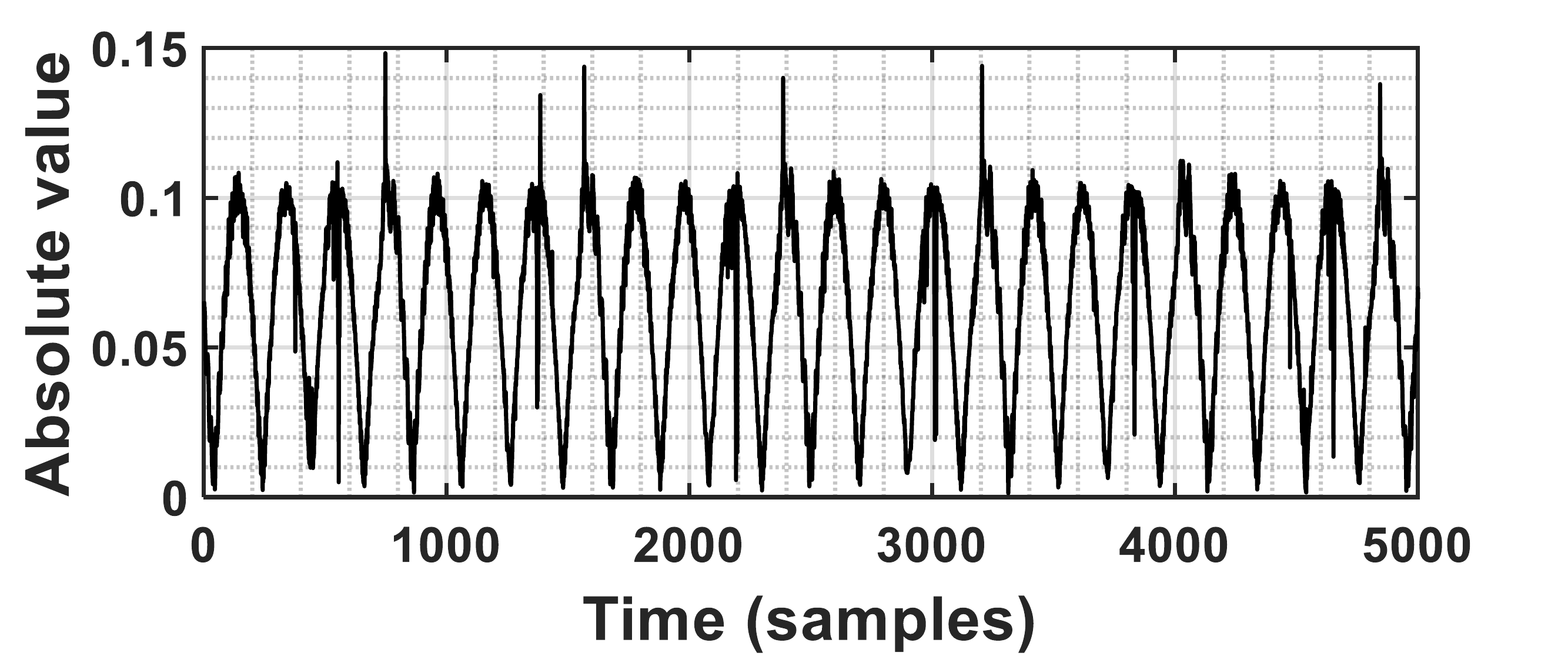}
\caption{50~$\mu$s time offset}
\end{subfigure}
\begin{subfigure}[b]{0.495\linewidth}
\includegraphics[width=1\textwidth]{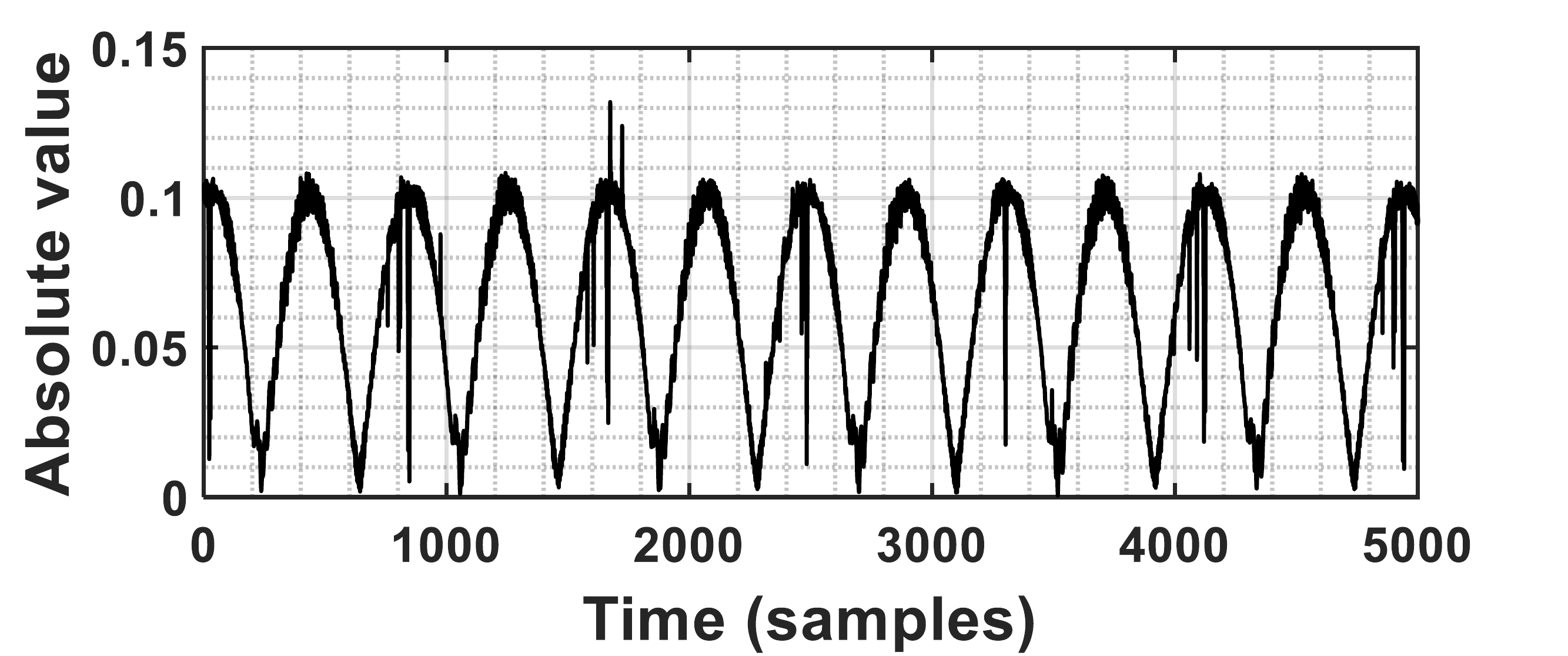}
\caption{25~$\mu$s time offset}
\end{subfigure}
\begin{subfigure}[b]{0.495\linewidth}
\includegraphics[width=1\textwidth]{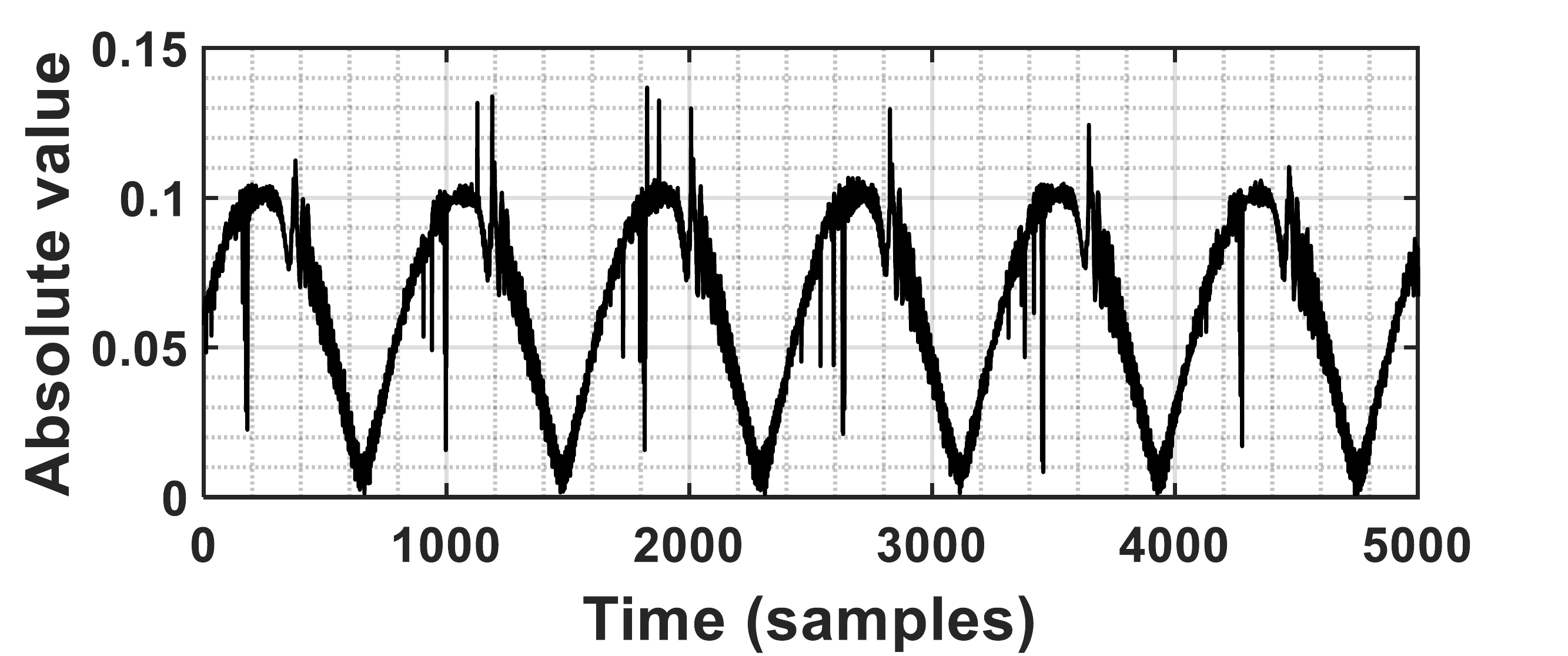}
\caption{12.5~$\mu$s time offset}
\end{subfigure}
\begin{subfigure}[b]{0.495\linewidth}
\includegraphics[width=1\textwidth]{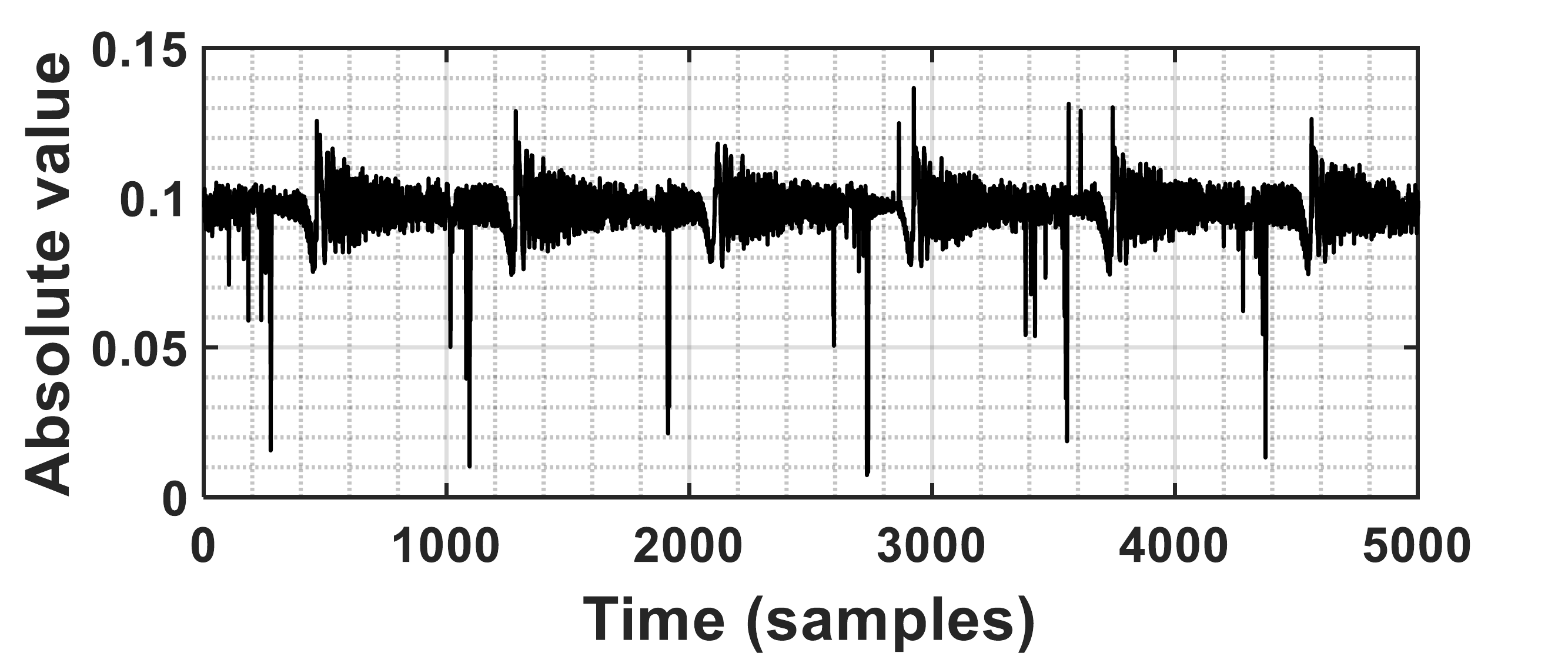}
\caption{0~$\mu$s time offset}
\end{subfigure}
\caption{\textbf{RSS measurements in different time offset settings.} A larger time offset leads to a higher RSS fluctuation rate. The RSS converges to a relatively stable value when all chirps are tightly time synchronized.}
\label{fig:sync_measure}
\end{figure}

To minimize the power consumption of the backscatter radio, we offload most of its operations to the radios outside the human body.
Taking a step further, we adopt chirp spread spectrum (CSS) -- a chirp pulse modulation that linearly sweeps a frequency band to generate the carrier signal -- to further cut down the power consumption.
Compared with conventional sinusoidal tone, CSS enables the wireless signal to be decodable below the noise floor (\eg, -137dBm for LoRa~\cite{lorasensitivity}) by introducing the unique processing gain on the frequency domain.
Given a fixed transmission power, the processing gain~($PG$) is proportional to the product of the chirp symbol time $S_t$ and the bandwidth $S_{bw}$: $PG \propto S_t\times S_{bw}$~\cite{olssonexploring}. 
We can thus have different trade-offs between system delay and  spectrum utilization in different scenarios, without hurting the signal detection accuracy.
In the following examples, we set the chirp bandwidth and symbol time to 40~KHz and 4~$ms$.
We have also explored other settings in our evaluation (\S\ref{exp_sss:beamforming}).

\paragraph{Chirp synchronization}. Tight time synchronization of chirp signals is the key to the success of beamforming.
Otherwise the beamforming power will fluctuate drastically due to the periodical coherent and incoherent signal combinations.
We design a \emph{two-step chirp synchronization algorithm} for this purpose. 
In the first step, the leader radio broadcasts a chirp preamble. 
The slave radio synchronizes with this preamble through cross-correlation.
The resulting lag then translates into a sample offset between the reference chirp and the received chirp preamble. 
Each slave radio can thus compensate for this initial time offset. 
However, due to heterogeneous software and hardware processing delays among radios, residual time offset still remains.

In the second step, the slave radios transmit a continuous chirp signal; the leader radio listens.
All slave radios then take turns to compensate for the residual time offset under the guidance of the leader radio. 
This is based on the realization that the amplitude of the superimposed signal (at the leader radio) will fluctuate periodically if the incoming chirps are not tightly time synchronized. 
In fact, the larger the time offset, the faster the received signal amplitude fluctuates, as shown in Figure~\ref{fig:sync_measure}.
The leader radio computes the fluctuation rate of the received signal amplitude using fast Fourier transform (FFT) and then guides slave radios to compensate for the residual time offset.

\begin{figure}[t]
     \centering
     \includegraphics[width=\columnwidth]{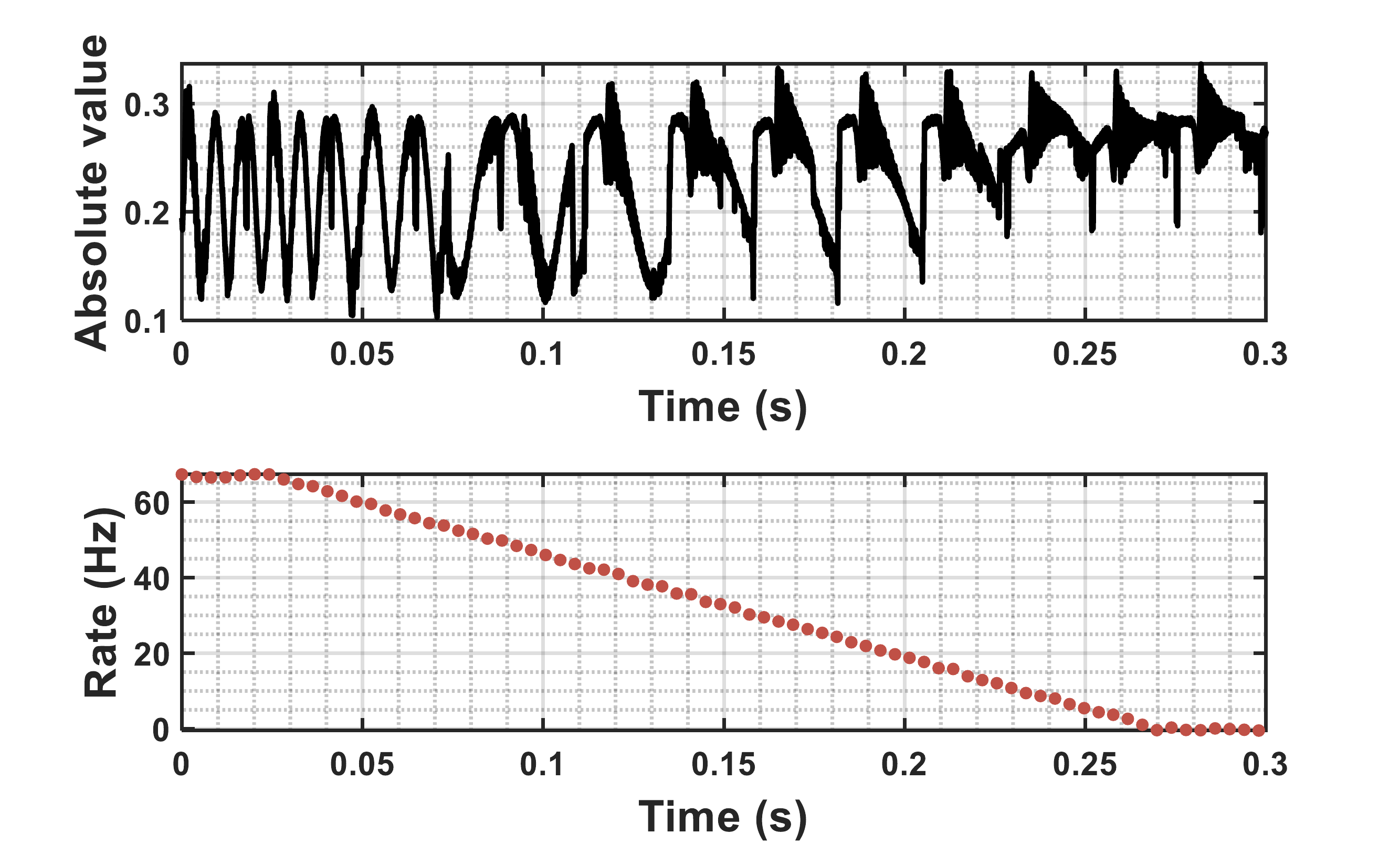}
     \caption{\textbf{A snapshot of RSS samples (top) and the fluctuation rate (bottom) in one period}. The fluctuation rate decreases with time. The signal amplitude converges when the two radios are synchronized. }
\label{fig:RSSsamples}
\end{figure}

The second step goes through a total of $N-1$ periods.
In each period $i$, \systemname aligns the initial time of the $i+1^{th}$ slave to the first slave.
Specifically, in the first period, two slave radios $S_1$ and $S_2$ send a continuous chirp signal simultaneously. 
These two signals add up at the leader radio.
Since $S_1$ and $S_2$ are not strictly time synchronized, we will see fluctuations of the received signal at the leader radio.
The leader then sends a two-bit feedback to $S_2$, telling this node to add or subtract one sample time, or to stop.
$S_2$ calibrates its clock based on this feedback, and then regenerate a chirp signal with an updated clock.
The leader radio detects the change of the fluctuation rate and sends an updated feedback to $S_2$.
The algorithm iterates as above until all slave radios are synchronized.
The algorithm then enters the next period and involves one more slave radio.
All slaves are tightly time synchronized at the end of the last period.

Considering its iterative nature, one may fear our synchronization algorithm may cause an excessively long delay.
However, the first step of the algorithm can already yield small residual time offset and usually a reasonable number of iterations (tens) are needed in each period. 
Figure~\ref{fig:RSSsamples} shows the variation of received signals (top) and the fluctuation rate (bottom) in one period.
The fluctuation rate drops to almost zero in 0.27~s.

\begin{figure}[t]
\centering
{\includegraphics[width=0.8\columnwidth]{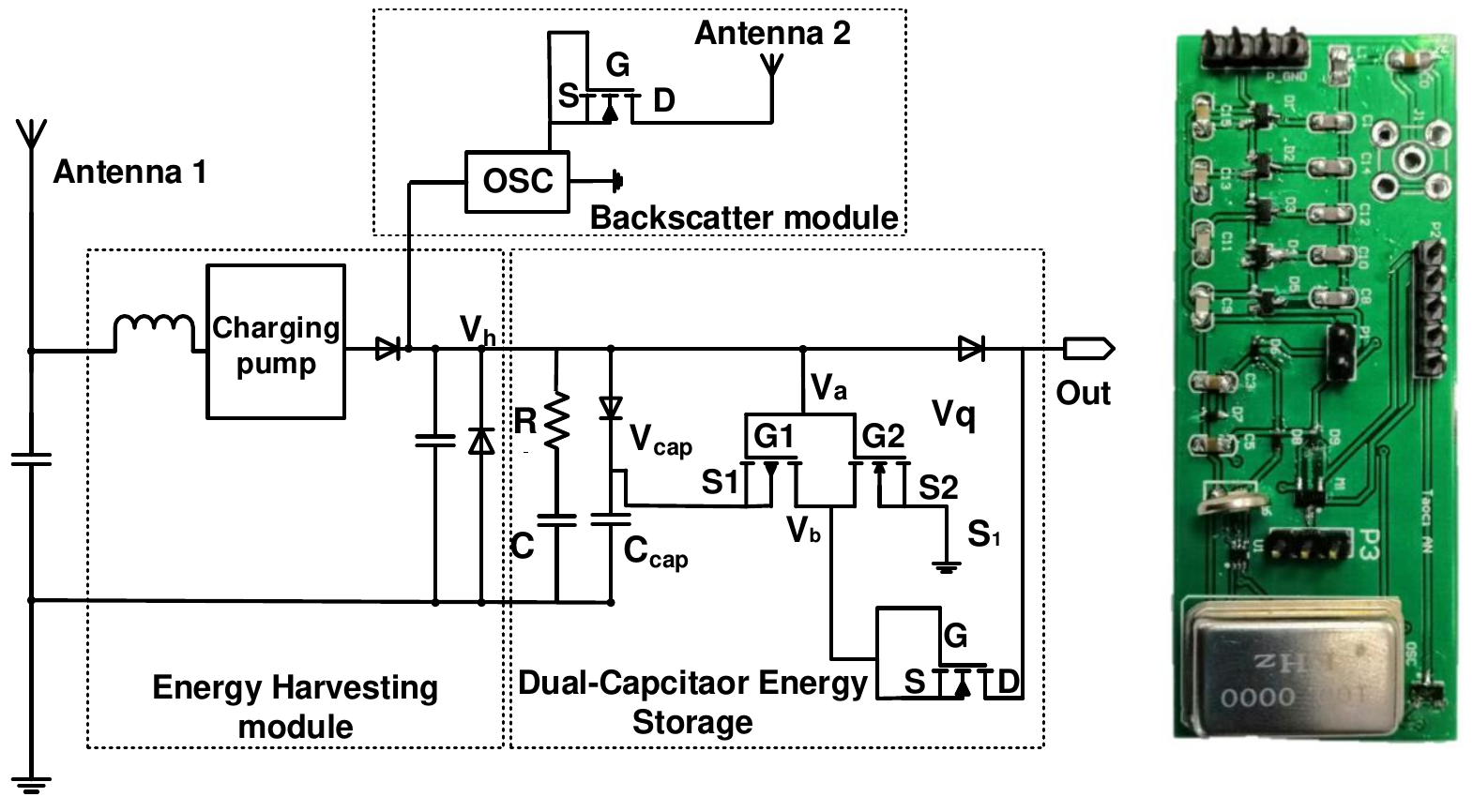}}
\caption{\textbf{Our monotonic backscatter radio design (left) and the PCB board prototype (right)}. 
We envision the board size can be reduced to the sub-centimeter scale when implementing \systemname on an Integrated Circuit (IC).}
\label{fig:backscatterRadio}
\end{figure}

\subsubsection{Backscatter Design}
\label{sss:backscatterRadio}

The chirp modulation enables the leader radio to detect the weak backscatter signal.
However, generating this backscatter signal requires the backscatter radio to measure the received power and compare it with the signal power measured in previous round.
These operations require extra hardware, computation and more importantly, power consumption.

To solve this challenge, we offload power measurement from the backscatter radio to the leader radio outside human body.
We choose this design based on the key observation of the monotonic backscatter system: the backscatter signal power chan\-ges monotonically with the received beamforming power.
By observing the power change of the received backs\-catter signals, the leader radio could infer the power change of the received beamforming power. 

In \systemname, the backscatter radio shifts the superimposed carrier signal to another frequency band (for interference avoidance) and reflects it directly back to the leader radio.
This is achieved by letting the backscatter radio generate a baseband signal at frequency $f_s$ and mix this baseband with the superimposed carrier signal at frequency $f_1$.
The mixer operation will shift this superimposed carrier signal to another two frequency bands: $f_1+f_s$ and $f_1 - f_s$.
The leader radio detects the backscatter signal on one of these two frequency bands and infers the beamforming power change accordingly.
Following the iterative beamforming algorithm introduced in \S\ref{ss:design_beamformingWithoutCSI}, the leader radio then guides slave radios to adjust their signal phase settings.
To avoid interference between the carrier signal and the backscatter signal, we conservatively set $f_s$ to 100~KHz, which is 1.5$\times$ larger than the default chirp bandwidth (40~KHz).

\begin{figure}[t]
\centering
{\includegraphics[width=1\columnwidth]{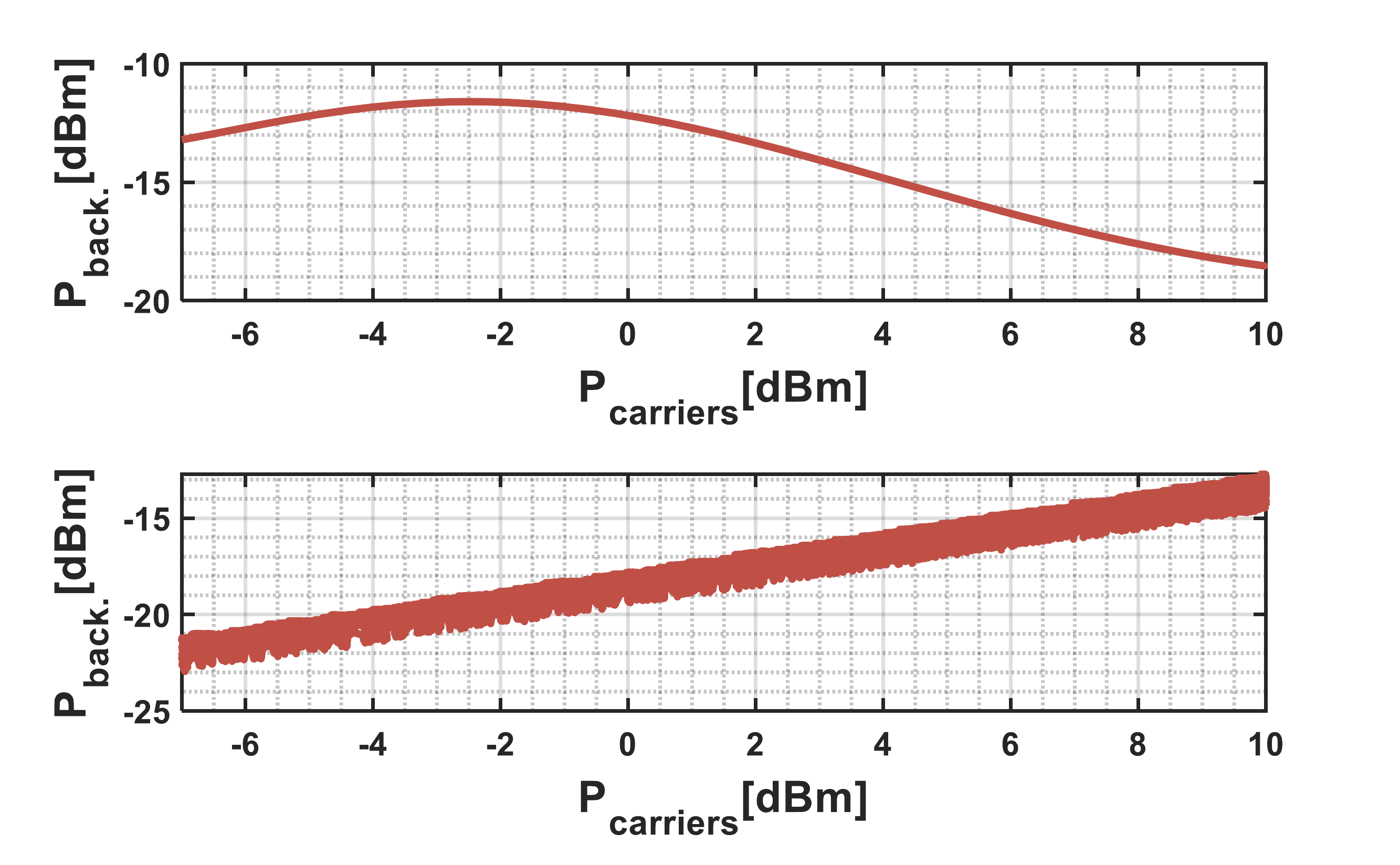}}
\caption{\textbf{Non-monotonic (top, passive RFID~\cite{arnitz2013wireless}) and monotonic backscatter radio (bottom, \systemname)}.
The top figure is adapted from \cite{arnitz2013wireless}.}
\label{fig:linearPower}
\end{figure}

\paragraph{Radio hardware design}. Conventional backscatter design (\eg, RFID), however, is not always monotonic, 
as shown in Figure~\ref{fig:linearPower}(top).
This non-monotonic property arises from the energy harvesting circuit where the impedance of matching network changes with input RF power.


To this end we design a low-power, monotonic backscatter radio. 
The hardware schematic is shown in Figure~\ref{fig:backscatterRadio}.
Our backscatter radio contains two RF chains, one for energy harvesting (through antenna one) and another for backscatter (through antenna two).
It allows the energy harvesting and backscatter to operate in parallel, without interfering each other.
To achieve a consistent impedance, the RF power on the backscatter radio should be relatively stable.
Hence we put a diode in-between these two modules, which allows the electric current to pass through in one direction (from the bacskcatter module to the energy harvesting module), while block it in the opposite direction.
We measure the backscatter signal power as we gradually increase the carrier signal power.
The result is shown in Figure~\ref{fig:linearPower}.
We observe that the backscatter signal power changes monotonically with the carrier signal, which confirms the effectiveness of the hardware design.
The dynamic power consumption of this backscatter radio is 42~$\mu$W, which takes up only around 12$\%$ of the energy harvested from our testbed (0.37~$mW$).


\subsubsection{Beamforming Power Change Inference}
\label{sss:powerTrendInference}

\begin{figure}[t]
\centering
\begin{subfigure}[t]{0.49\linewidth}
\centering
\includegraphics[width=1.1\textwidth]{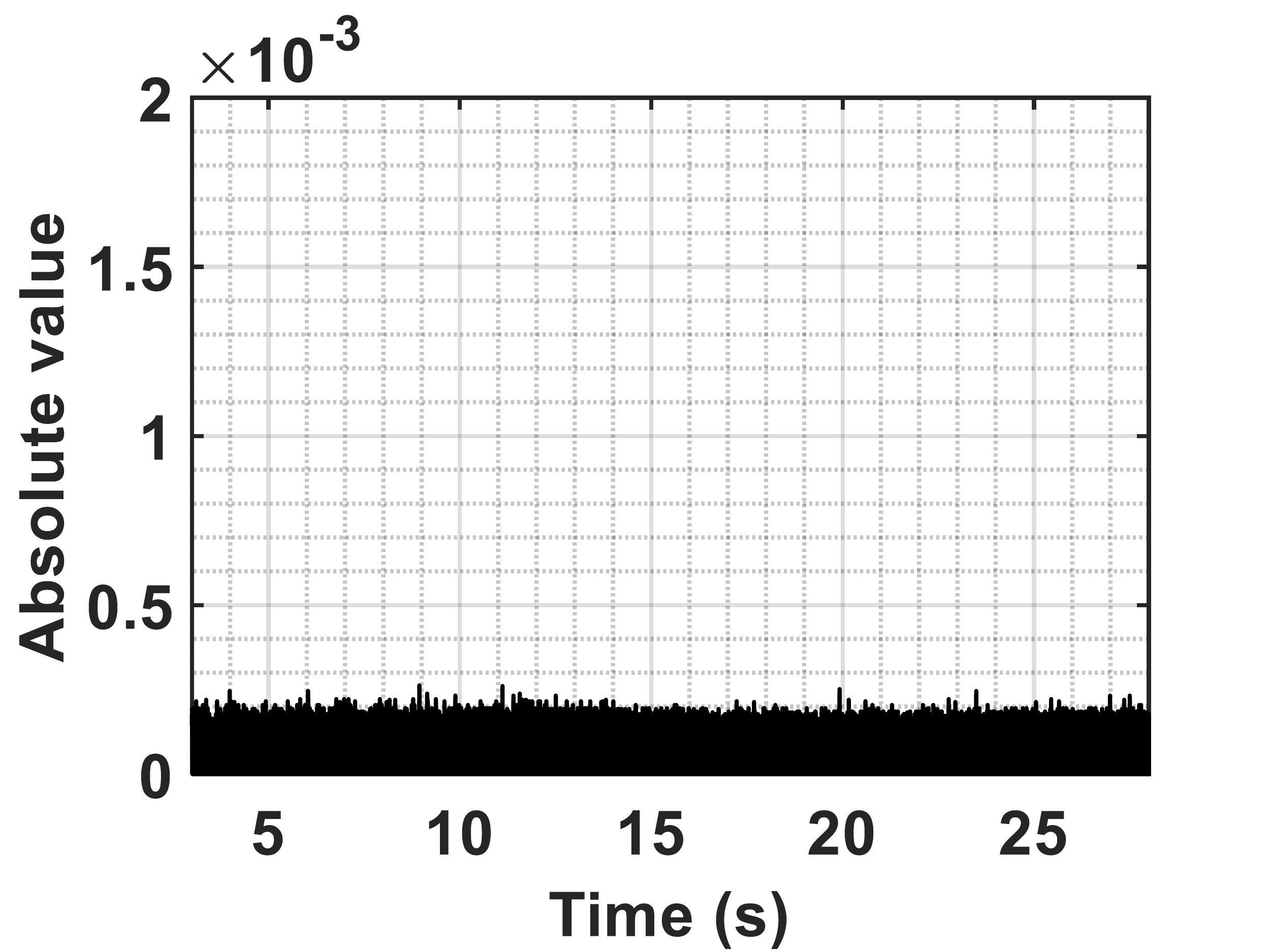}
\caption{RSS measurements}
\label{fig:example_ccs}
\end{subfigure}
\begin{subfigure}[t]{0.49\linewidth}
\centering
{\includegraphics[width=1.1\linewidth]{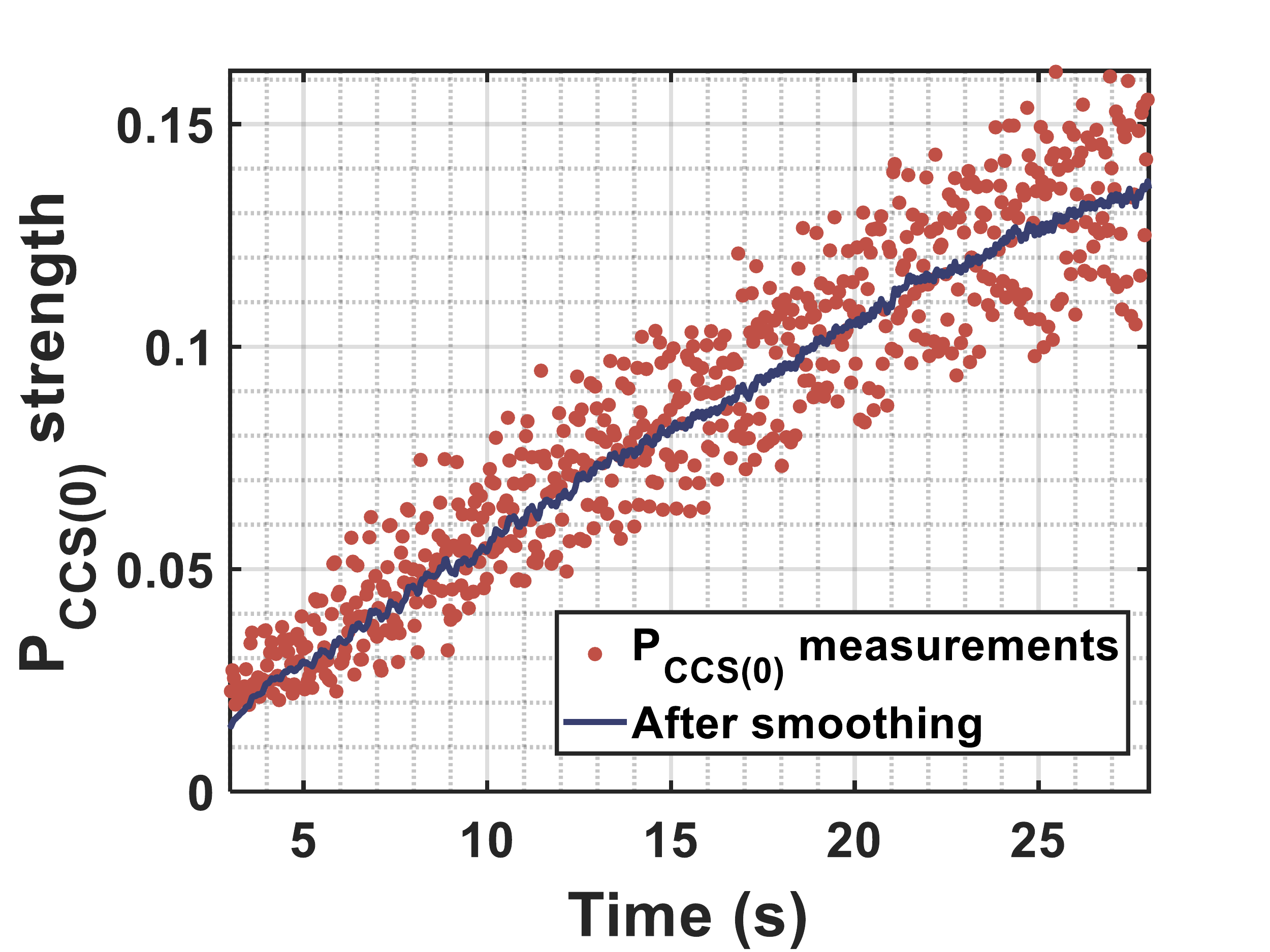}}
\caption{P$_{CCS(0)}$ measurements}
\label{fig:kf}
\end{subfigure}
\caption{\textbf{(a) RSS and (b) $P_{CCS(0)}$ measurements of the received backscatter signal as the carrier signal power grows linearly}.}
\end{figure}

The leader radio infers the power change of the beamforming signal by observing the power change of the received backscatter signal.
A new challenge arises due to the extremely weak backscatter signal  --  after going through considerable channel fading, the backscatter signal is  usually below the minimum detectable strength (MDS) of the commercial RF radios (\eg, around -75~dBm for an USRP N210 software defined radio~\cite{mds}).
To verify this challenge, we put a backscatter radio into a 10~cm thick pork belly and conduct the following experiment.
A transmitter node that is five meters away sends a continuous chirp pulse, with its power grows linearly from 0 to 20~dBm.
A receiver node that is one meter away from the backscatter radio measures the received backscatter signal.
Figure~\ref{fig:example_ccs} plots the amplitude of the received backscatter signal.
We observe noisy power measurements which fail to reflect the power change of the backscatter signal.

We define a new metric called \emph{$P_{CCS(\omega)}$} and use it to infer the power change of the backscatter signal.
\emph{$P_{CCS(\omega)}$} is computed by correlating the received backscatter signal with the reference chirp in the frequency domain.
We have proved that the peak value of \emph{$P_{CCS(\omega)}$}, namely, $P_{CCS(0)}$, changes monotonically with the power change of the backscatter signal, and demonstrated that $P_{CCS(0)}$ has sufficient resolution to reflect the power change of the backscatter signal.
Due to the page limitation, we put the mathematical proof in Appendix~\ref{s:p_ccs0}.

Figure~\ref{fig:kf} shows $P_{CCS(0)}$ samples extracted from the received signals. 
We observe an increasing trend of $P_{CCS(0)}$ as we increase the power of carrier signal.
However, due to signal noises and measurement errors, $P_{CCS(0)}$ fluctuates drastically, which may confuse the leader radio and introduce extra beamforming iterations.
To solve this problem, we adopt an adaptive Kalman filter~\cite{brown1992introduction} to smooth the $P_{CCS(0)}$ samples.
Figure~\ref{fig:kf} shows that the filtered samples can fairly reflect the power change of the backscatter signal.

\subsection{Cold Start}
\label{ss:coldStart}

Previous sections focus on how to beamform towards the backscatter radio that is already awake.
In this section, we describe how we bootstrap the backscatter radio during the cold start period. Cold start is a ``chicken-n-egg'' problem: without enough power (-20~dBm at least~\cite{R6}) the backscatter radio cannot wake up to provide feedback (by simply reflecting the signal).
On the other hand, without the feedback, we cannot beamform to provide energy. 
Exhaustively searching all the beamforming space in hope of accidentally waking up the backscatter radio is obviously not a viable approach. 
Employing PushID~\cite{pushid} to wake up the backscatter radio, on the other hand, requires a much stronger carrier signal to compensate the excessive channel fading inside the human body, which may overheating human tissues and cause safety issues.

\begin{figure}[t]
\centering
\begin{subfigure}[b]{0.49\linewidth}
\includegraphics[width=1.1\textwidth]{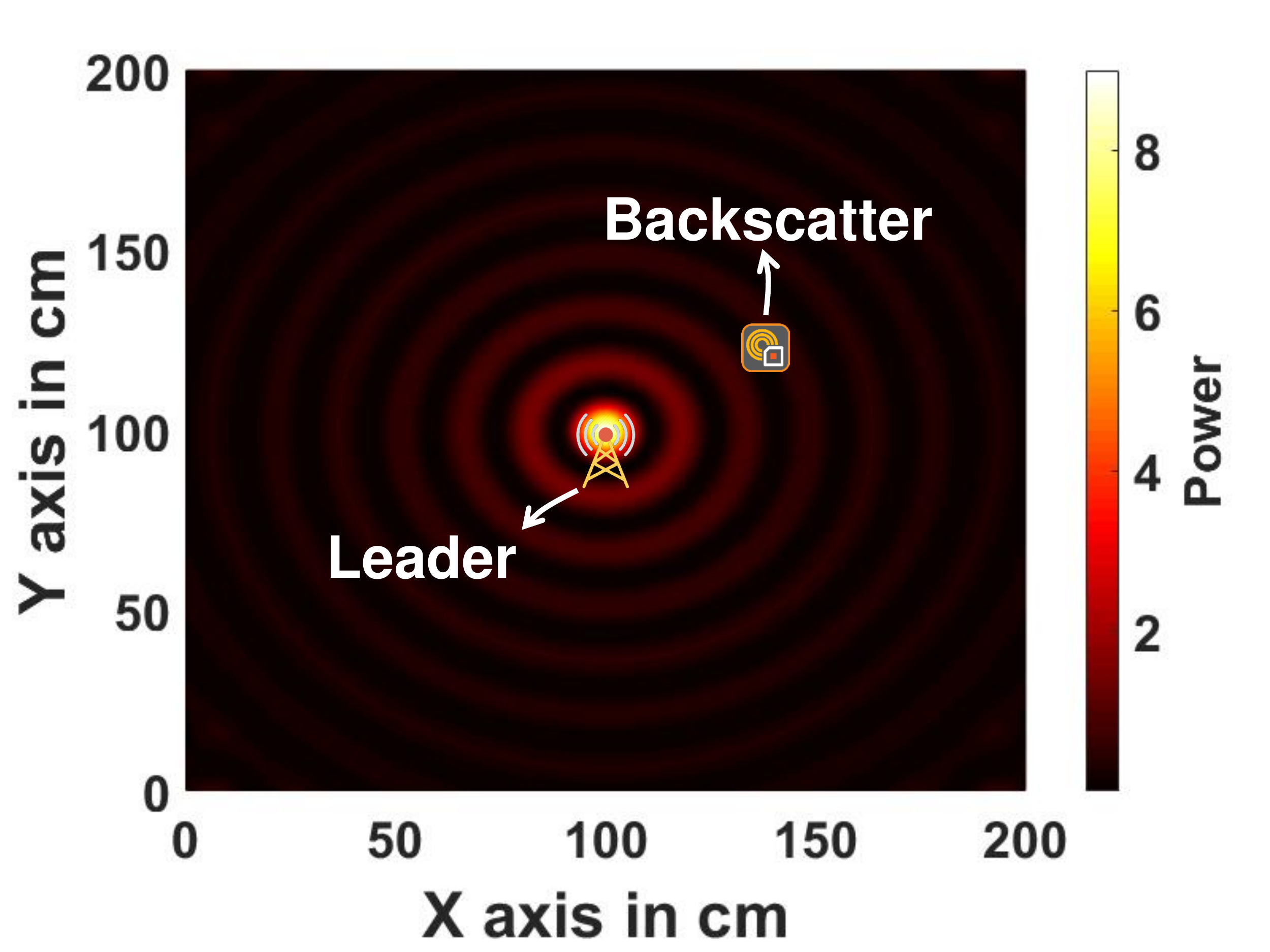}
\caption{}
\end{subfigure}
\begin{subfigure}[b]{0.49\linewidth}
\includegraphics[width=1.1\textwidth]{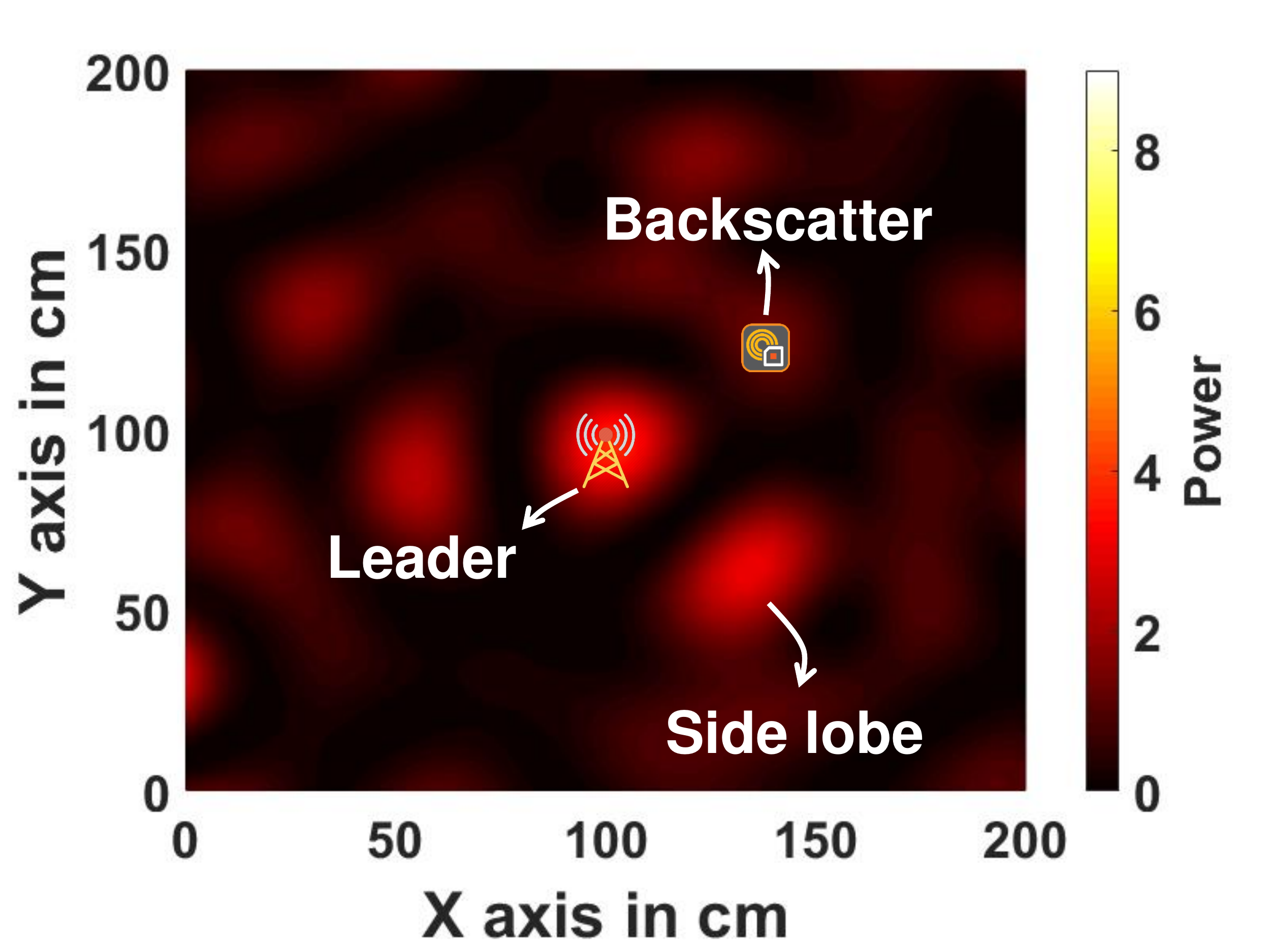}
\caption{}
\end{subfigure}
\begin{subfigure}[b]{0.49\linewidth}
\includegraphics[width=1.1\textwidth]{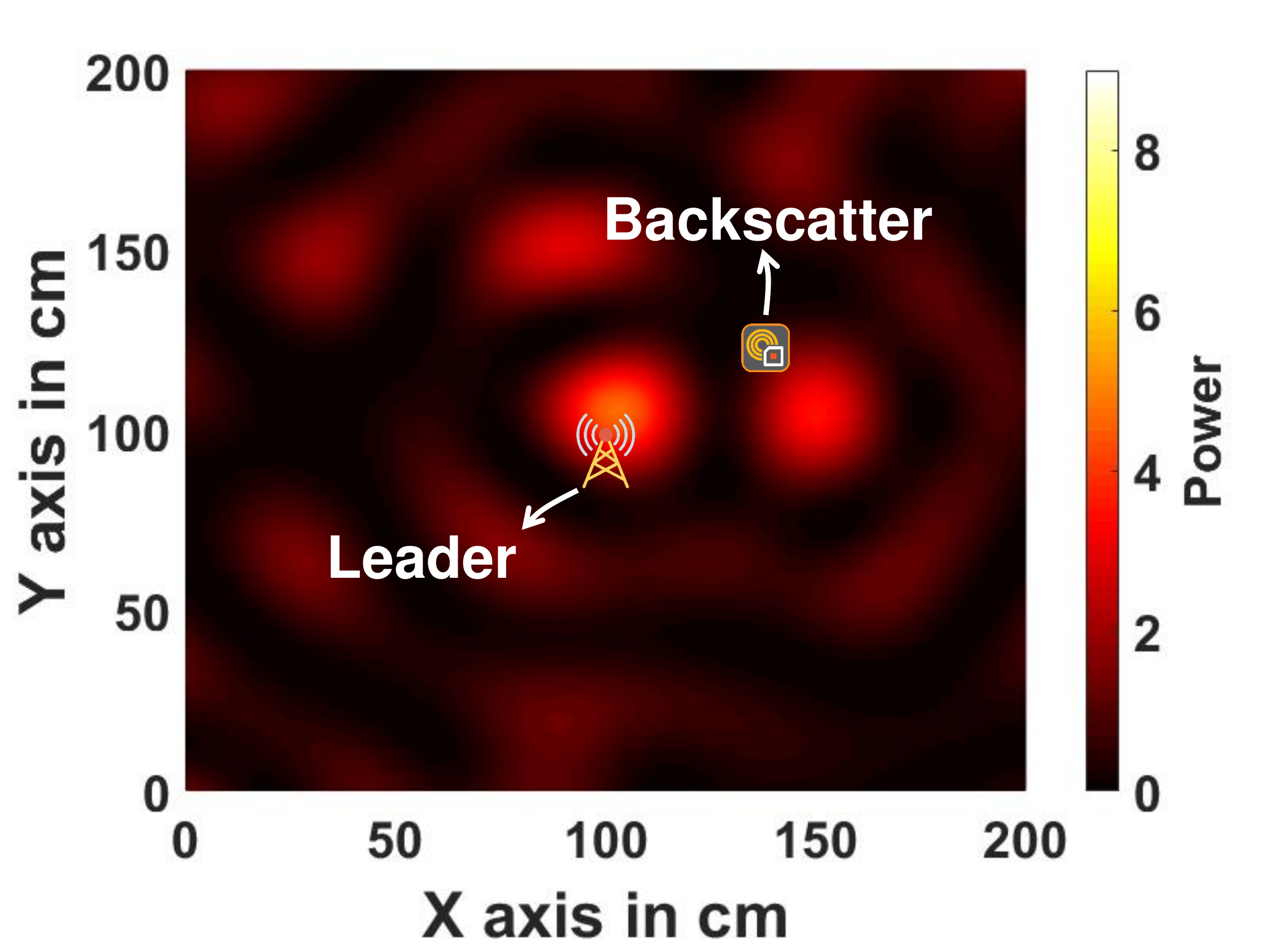}
\caption{}
\end{subfigure}
\begin{subfigure}[b]{0.49\linewidth}
\includegraphics[width=1.1\textwidth]{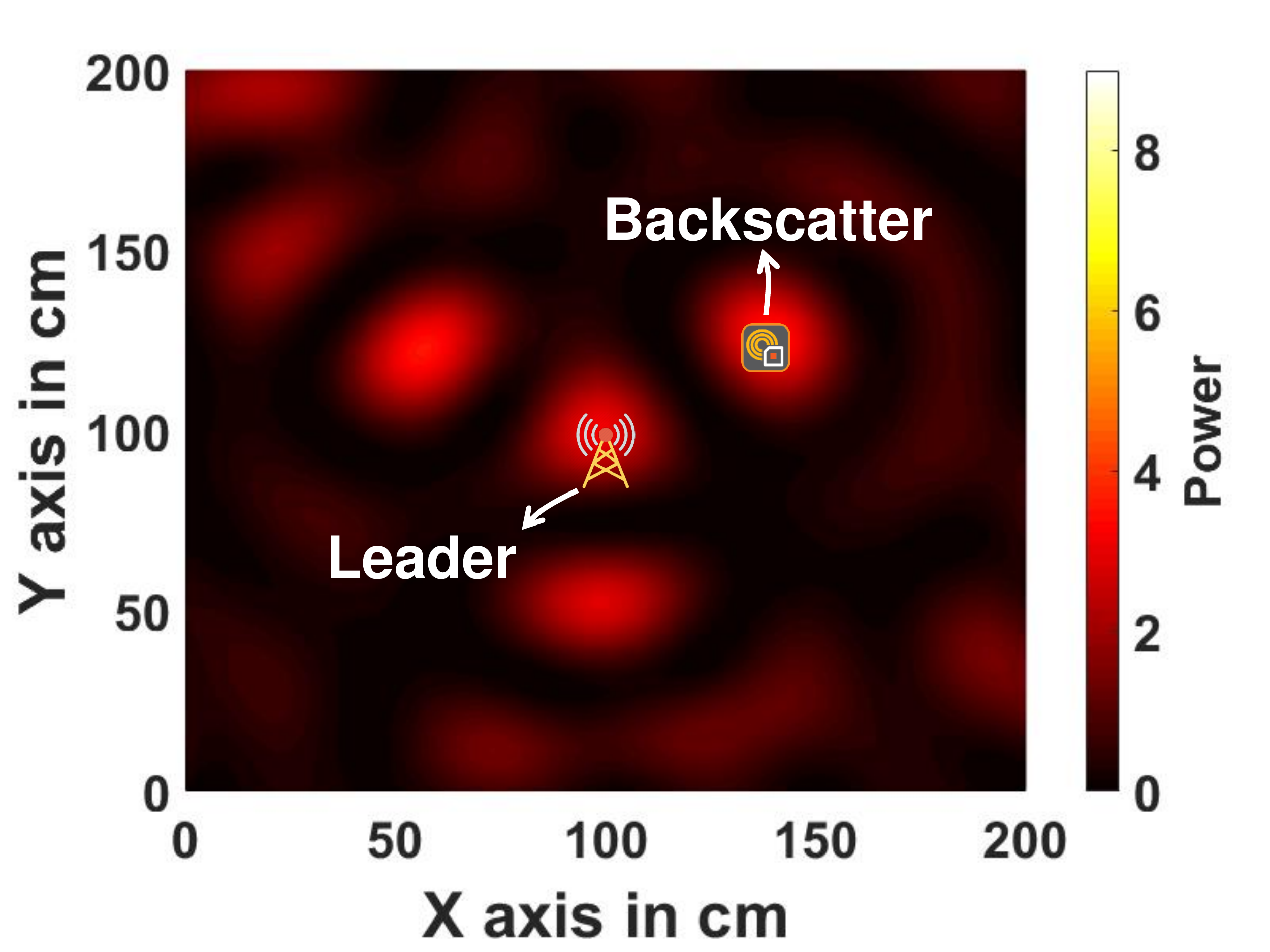}
\caption{}
\end{subfigure}
\caption{\textbf{Beamforming energy patterns with different phase perturbations}:
(a) optimal phase alignment, (b)-(d) with different phase perturbations.
With phase perturbations, we observe an enlarged main lobe and many side lobes.
These side lobes can provide sufficient energy to wake up the backscatter radio.}
\label{fig:cs}
\end{figure}

We propose a beamforming-based space searching algorithm to bootstrap the backscatter radio. 
Recall that the leader node can be a mobile phone or a wearable device worn by the user, it is thus reasonable to assume the leader node is close to the medical implant.
In \systemname, we first align all beams towards the leader node and then search the limited space around the leader node.
The space searching algorithm is based on the realization that different phase combining can lead to a significant different beamforming patterns.
Specifically, let $\phi_{i}$ be the current phase setting of the slave radio $i$.
As we introduce a phase perturbation $\delta_{\phi}$ to $\phi_{i}$, the carrier signals will coherently combine at other locations, resulting in side lobes.
This new phase combination also spreads the main beamforming lobe over a larger area, as shown in Figure~\ref{fig:cs}(b).
Accordingly, by introducing different phase perturbation $\delta_{\phi}$~( $-\sigma<\delta_{\phi}<\sigma$) to each slave radio, we can produce different beamforming patterns and use them for space searching.
When the backscatter radio gains enough energy as a result of this searching effort, it wakes up and starts to backscatter. 
Once the leader radio receives this bacskcatter signal, it goes back to serve its functions described in \S\ref{sss:carrierSignal} and \S\ref{sss:powerTrendInference}.

As the beamforming power spreads over the main lobe and side lobes, the question here is whether these lobes are strong enough to wake up the backscatter radio.
To answer this question, we measure the power distribution of the beams shown in Figure~\ref{fig:cs}(b)--(d) and find a 3.6~dB power drop with repesect to the optimal beamforming power (Figure~\ref{fig:cs}(a)).
Note that to achieve a desirable charging efficiency, the optimal beamforming power of a multi-antenna system is much higher than the power required in the cold start period (-15~dBm). Hence, these newly emerging beams are strong enough to wake up the backscatter radio.
Our micro-benchmark result (\S\ref{exp_sss:coldStart}) also confirms the efficacy of this cold start method. 

\begin{figure}[t]
\centering
\begin{subfigure}[b]{0.49\linewidth}
\includegraphics[width=1.1\textwidth]{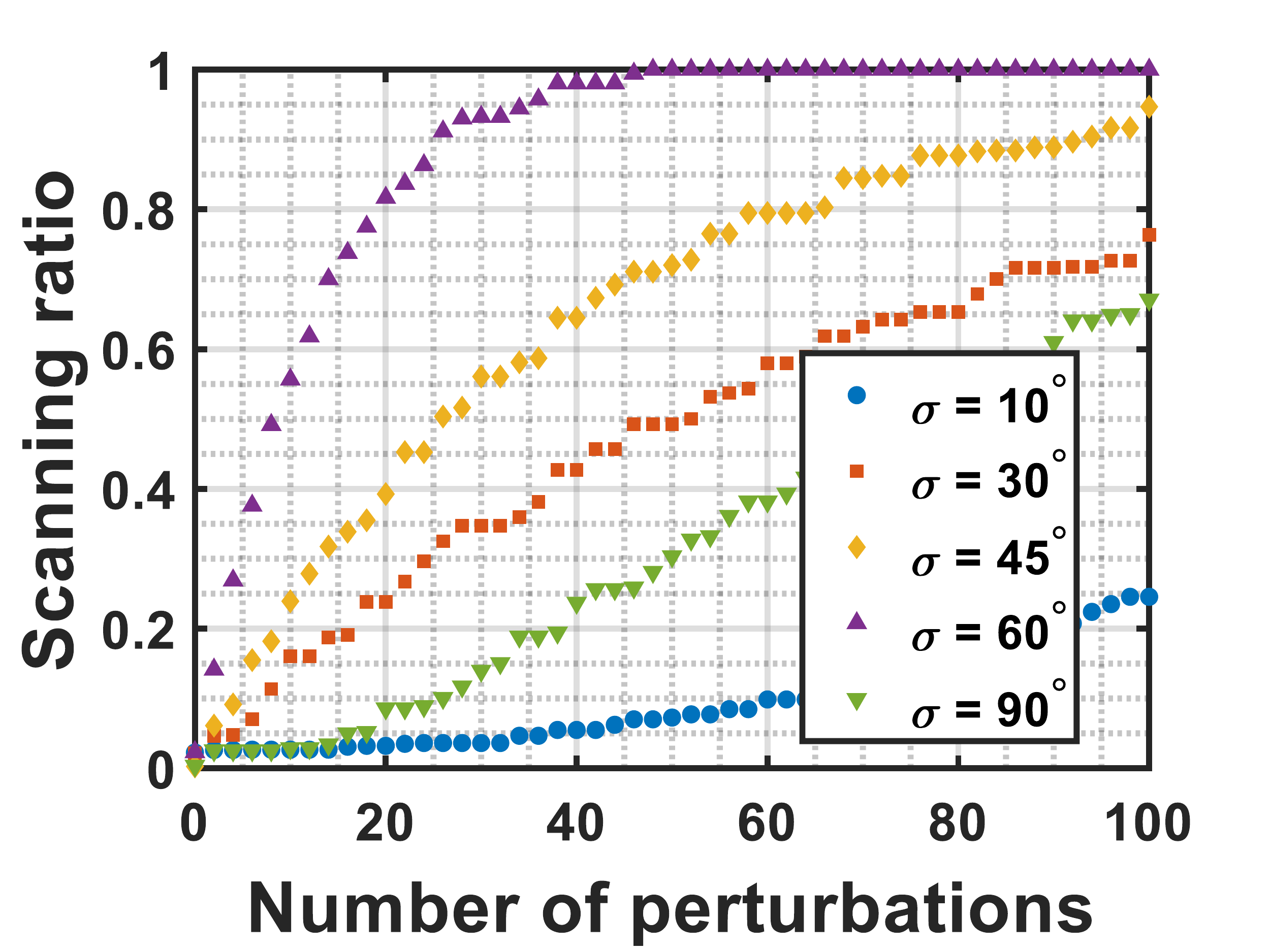}
\caption{Scanning ratio CDF vs. number of perturbations}
\end{subfigure}
\begin{subfigure}[b]{0.49\linewidth}
\includegraphics[width=1.1\textwidth]{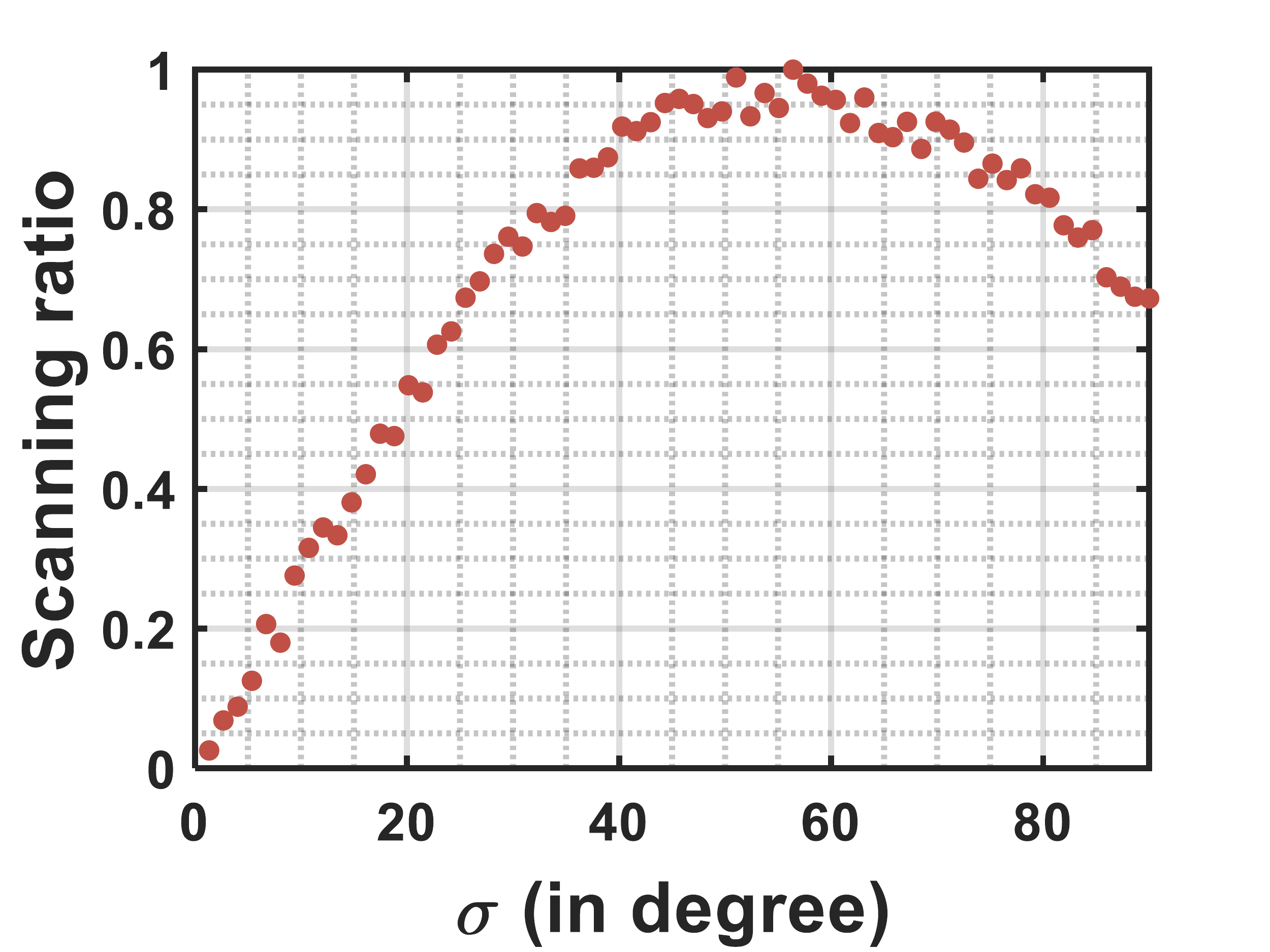}
\caption{Scanning ratio values after 100 perturbations vs. $\sigma$}
\end{subfigure}
\caption{\textbf{Phase perturbation range $\sigma$ analysis}.}
\label{fig:css_para}
\end{figure}

\paragraph{Determining the phase perturbation range $\sigma$}. 
We define the scanned area as the space where the received energy is higher than $30\%$ of the optimal beamforming power (equivalent to  < 5~dB loss).
We then conduct simulations to investigate the impact of the phase perturbation range $\sigma$ on the scanning ratio -- the ratio of the scanned area to the entire searching space (a 2$\times$2$\times$2~m$^3$ Cube centered at the receiver).
Figure~\ref{fig:css_para}(a) shows the scanning ratio as a function of phase adjustment in different $\sigma$ settings. 
The scanning ratio grows rapidly as we increase $\sigma$ from 10$^{\circ}$ to 30$^{\circ}$ and further to 60$^{\circ}$.
The growth of scanning ratio then slows down as $\sigma$ increases further.
To better understand this result, we further repeat the above experiment 100 times in different $\sigma$ settings and show the result in Figure~\ref{fig:css_para}(b).
We can see the scanning ratio peaks the maximal when $45^{\circ} \leq \sigma \leq 65^{\circ}$.
Suggested by this simulation result, we set $\sigma$ to 55$^\circ$.

\subsection{Balancing Convergence and Delay}
\label{sss:optPhaseAdjust}

In our iterative beamforming algorithm, the phase searching bound $\Phi$ is critical to system performance ($\Phi$ is introduced in \S\ref{ss:design_beamformingWithoutCSI}). 
If $\Phi$ is too large, the algorithm may rapidly converge to a non-optimal beamforming result.
In contrast, a smaller $\Phi$ will lead to better beamforming results, but with a longer delay. 
In \systemname, we use a larger phase bound at the beginning of the algorithm and then a smaller value as the algorithm iterates.
We compute a suitable phase searching bound in each iteration based on a high order polynomial function $\Phi=P(n)$, where $n$ is the iteration index.
Due to page limitation, we detail this polynomial function and its derivation in Appendix~\ref{sec:OPS}.

\begin{figure}[t]
\centering
{\includegraphics[width=1\columnwidth]{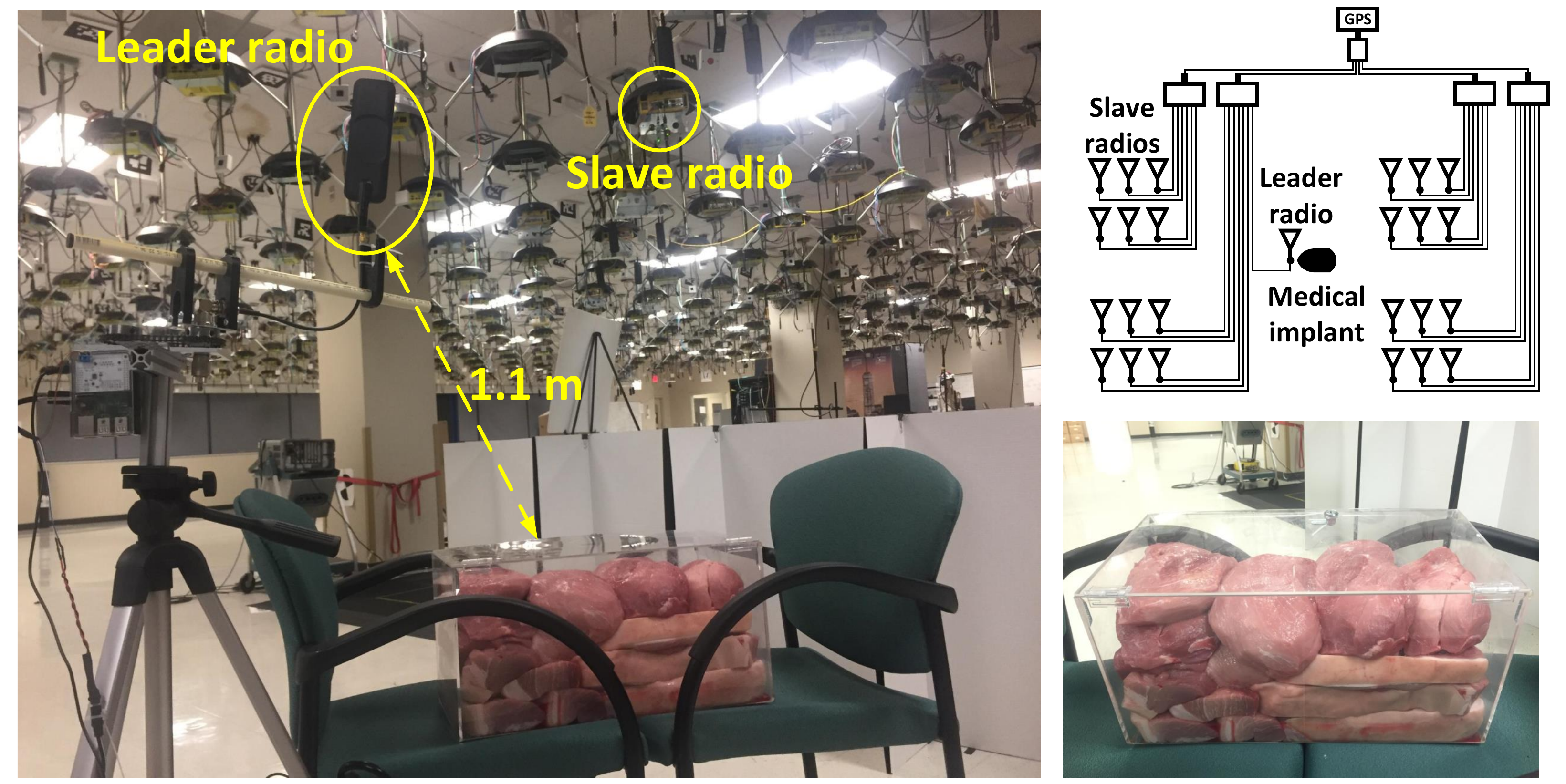}}
\caption{\textbf{Testbed setup}. Our testbed consists of 17 USRP N$210$ and four USRP B$210$ nodes, all mounted on the ceiling of an office building.}
\label{fig:exSetup}
\end{figure}

\begin{figure}[t]
\centering
{\includegraphics[width=1\columnwidth]{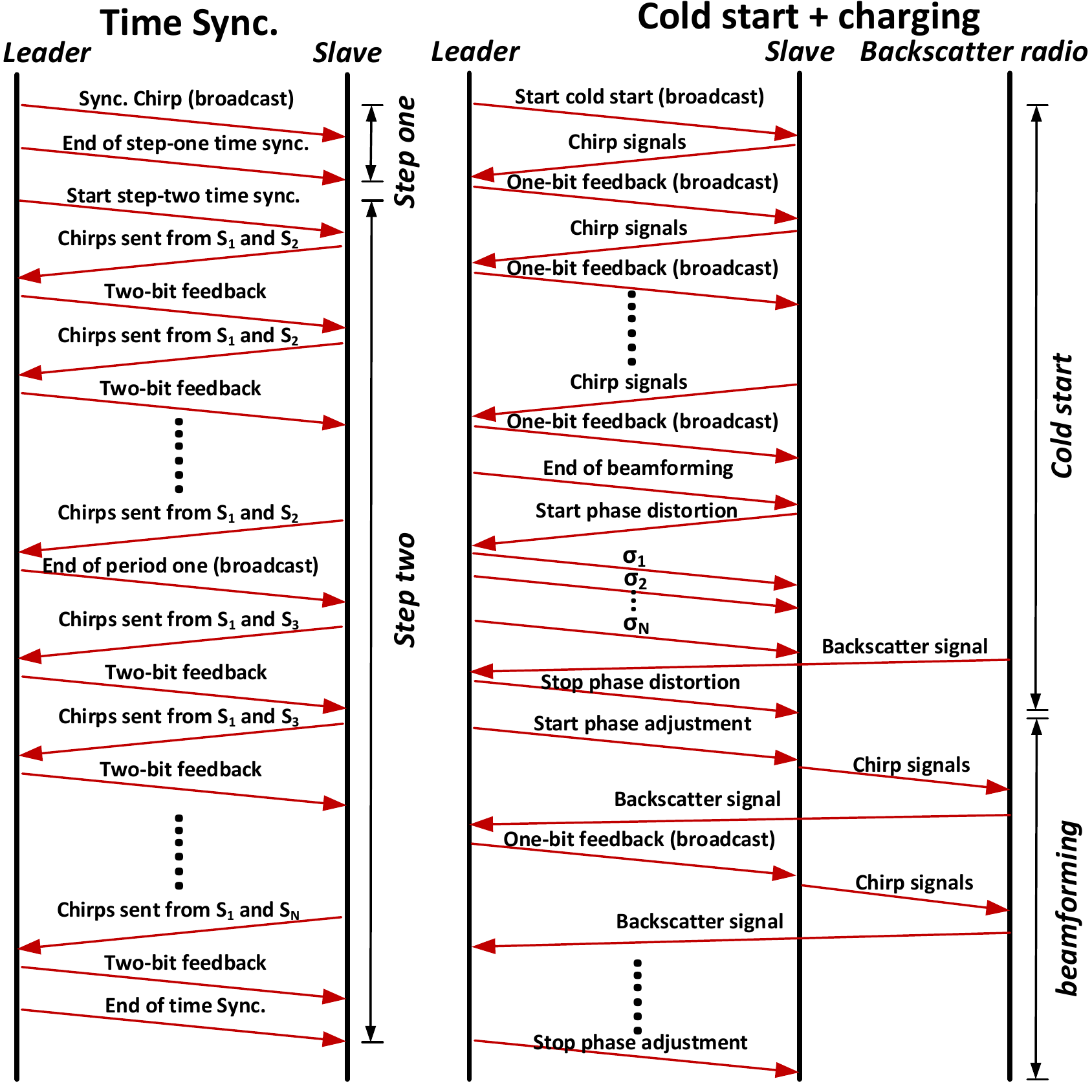}}
\caption{\textbf{Message flow of \systemname}.}
\label{fig:msgFlow}
\end{figure}

\section{Implementation}\label{s:implementation}

We describe the system implementation in this section.

\subsection{Testbed Setup}
We deploy 17 USRP N$210$ and four USRP B$210$ software defined radios on the ceiling of an office building, 
as illustrated in Figure~\ref{fig:exSetup}.
Each USRP is equipped with a WBX RF daughter board~\cite{dboard} and works on FDD full duplex mode.
We use a Mini\hyp{}Circuits ZFL\hyp{}1000VH RF amplifier~\cite{amp} to boost the signal power and send out the amplified signal through a 4~dBi Taoglas TG.$35$.$8113$ antenna~\cite{ant}.
As USRP only supports relative signal power measurement~\cite{linear}, we conduct a one-time power calibration using an Agilent E$4405$B spectrum analyzer~\cite{spectrum} to acquire the absolute signal power.

\paragraph{USRP Synchronization}. To mitigate the clock drift and carrier frequency offset (CFO), all USRPs are wired to an Oct\-oclock-G GPS disciplined oscillator (GP\-SDO)~\cite{clk} with $10$MHz reference signal. 
This centralized time synchronization method provides an accurate timing reference. 
Wirel\-ess-based time synchronization methods such as~\cite{hamed2018chorus,rahul2012jmb} can be further employed for an even larger system deployment.

\subsection{Software Implementation}
\label{ss:softwareImple}

We implement all signal processing modules in C++ (version 4.8.4) with UHD driver V3.10.1 and GNU Radio Companion V3.7.6.1. 
Figure~\ref{fig:msgFlow} shows the message flow of these signal processing modules.
We next describe the module implementation on the leader radio and the slave radio.

\paragraph{Leader radio} has three modules: chirp signal synchronization, backscatter radio cold start, and beamforming orchestration. 
We implement the following signal processing functions to support the above three modules: chirp preamble generation and transmission, RSS fluctuation detection, two-bit feedback signal generation and transmission, backscatter signal detection, $P_{CCS(0)}$ calculation, smoothing, and comparison.

\paragraph{Slave radios} participate in all the three modules mentioned above. 
We implement the following signal processing functions on each slave radio: chirp preamble detection, two-bit feedback signal detection and decoding, time calibration, beamforming signal detection, random number generator, phase adjustment, chirp carrier generation, and transmission.

\begin{figure*}[t]
   \begin{minipage}[t]{0.24\textwidth}
     \centering
     \includegraphics[width=1.05\linewidth]{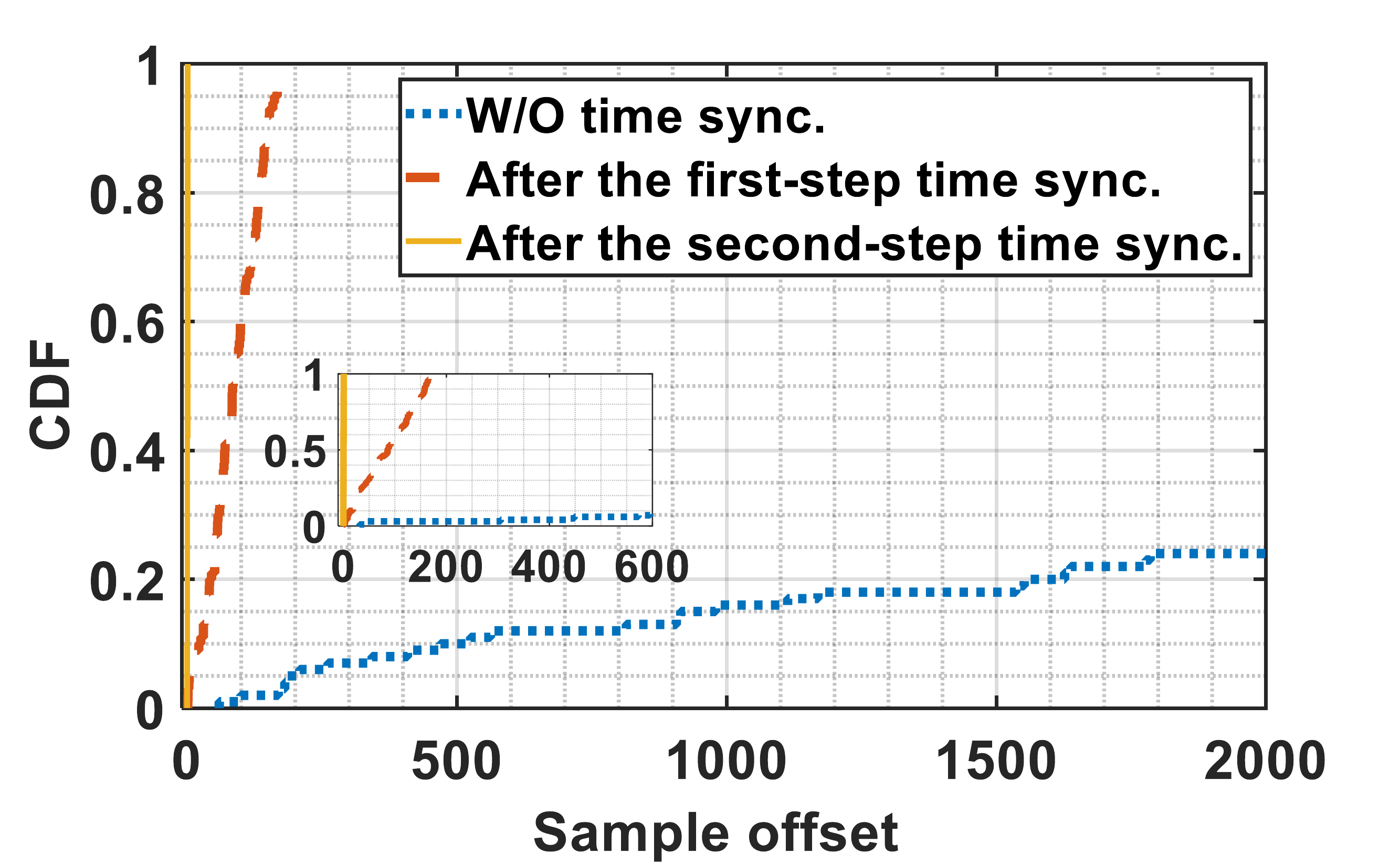}
\caption{CDF of residual time offset without and with chirp synchronization.}
\label{fig:snc1}
   \end{minipage}\hfill
   \begin{minipage}[t]{0.24\textwidth}
     \centering
     \includegraphics[width=1.06\linewidth]{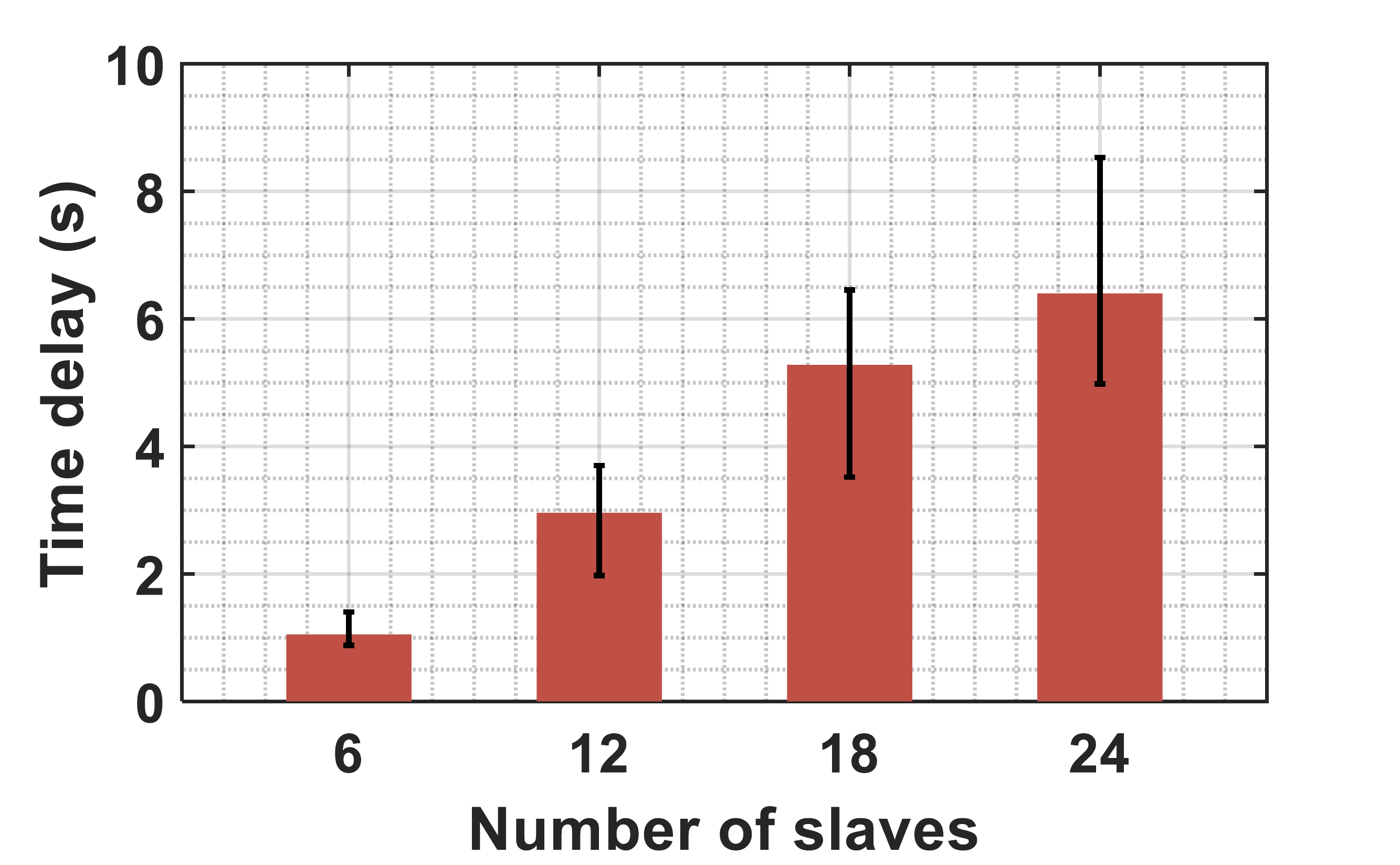}
\caption{Chirp synchronization delay vs. number of slave radios.}
\label{fig:snc2}
   \end{minipage}\hfill
   \begin{minipage}[t]{0.24\textwidth}
     \centering
{\includegraphics[width=1.05\linewidth]{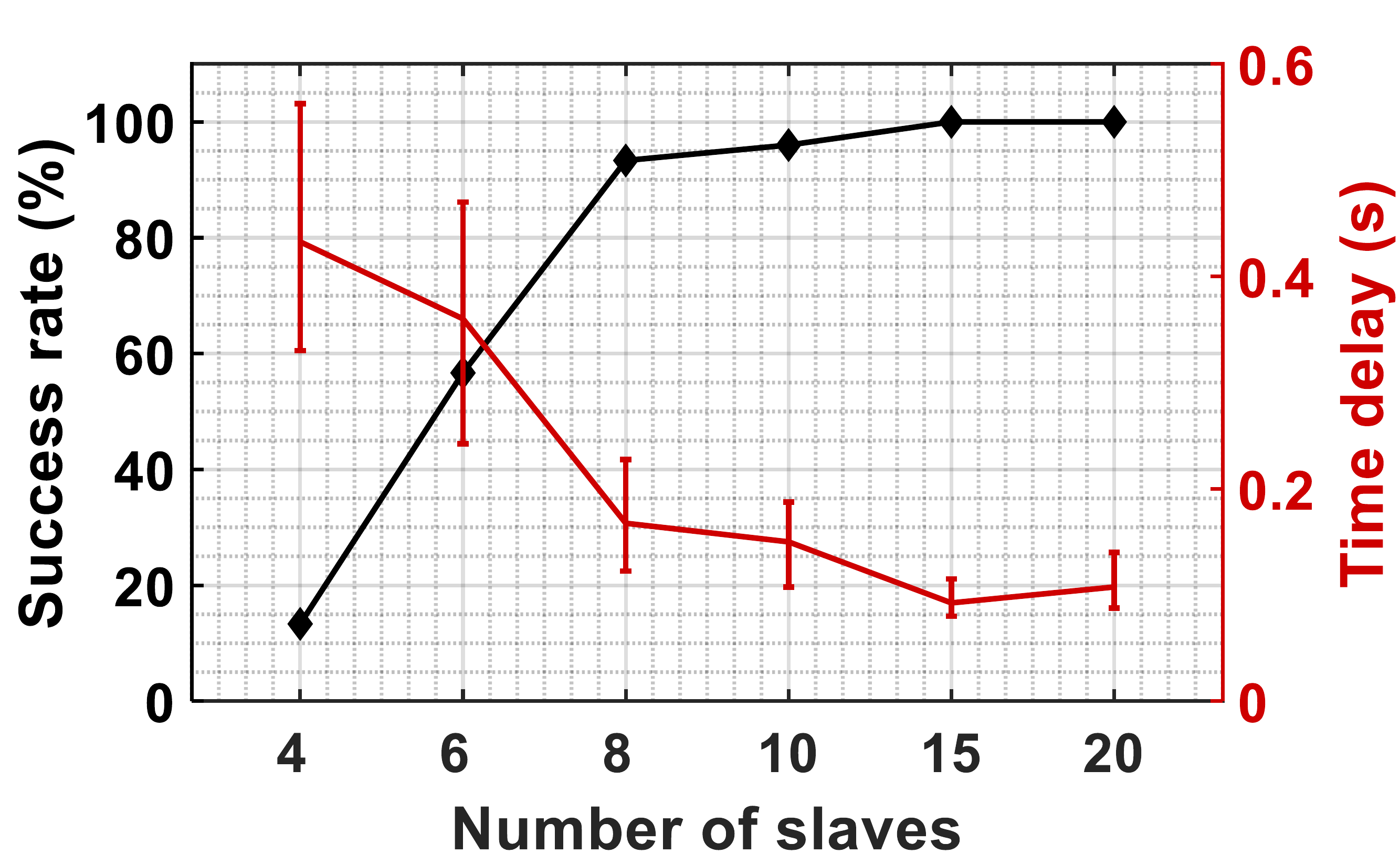}}
\caption{Cold start success rate (left) and delay (right) vs. number of slave radios.}
\label{fig:cs1}
   \end{minipage}\hfill
\begin{minipage}[t]{0.24\textwidth}
     \centering
    {\includegraphics[width=1.05\linewidth]{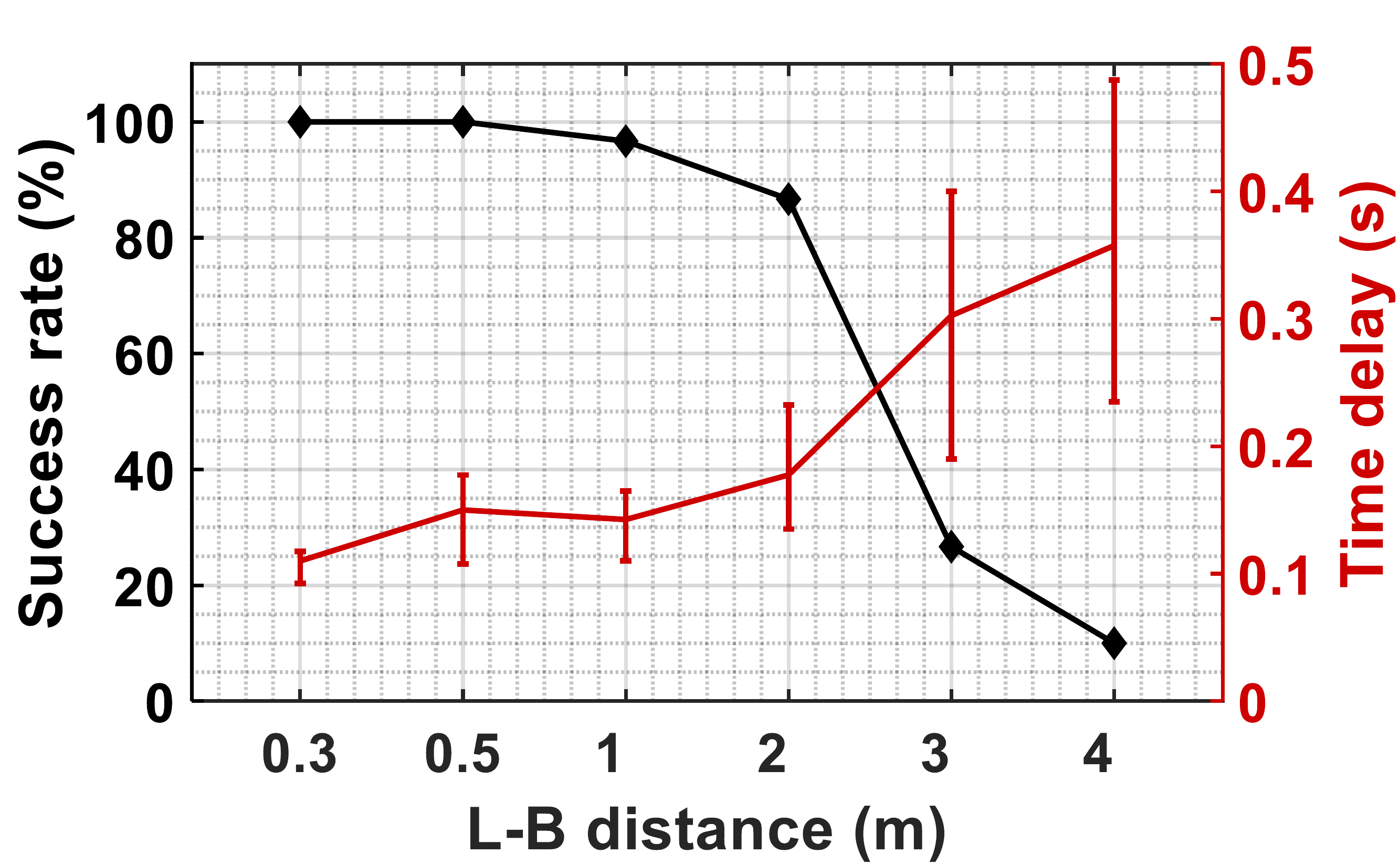}}
    \caption{Cold start success rate (left) and delay (right) vs. L--B distance.}
    \label{fig:cs2}
    \end{minipage}\hfill
\end{figure*}



\subsection{RF Power Limit}

Exposure to high levels of RF radiation can be harmful.

\paragraph{Transmission power:} According to Federal Communications Commission (FCC) regulation, the transmission power of a single radio (with a 4~dBi antenna gain) should be below 32~dBm~\cite{maxp}. 
In our deployment, the maximum transmission power is 30~dBm and thus complies with the FCC regulation.

\paragraph{Power density in space:} FCC and Food and Drug Administration (FDA) have different regulations  for power density. Specifically, FCC requires the power density in ISM band to be below 0.6~$mW/cm^2$~\cite{fcc}, whereas FDA requires the power density to be below 10~$mW/cm^2$~\cite{fda}. In our testbed, 24 antennas are distributed on the ceiling of an 18$\times$18~$m^2$ office building.
The theoretically maximal power density at the receiver is 0.08~$mW/cm^2$ based on~\cite{rfdensity}, which satisfies both FCC and FDA requirements. In our experiments, we also measure the beamforming power at different locations across the room. The maximum measured power density is  0.05~$mW/cm^2$, well below the power limits specified by FCC and FDA.

\section{Evaluation}
\label{s:evaluation}

We present the evaluation results in this section. 
By default the chirp bandwidth and symbol time are set as 40~KHz and 4~$ms$ (8192 samples), unless otherwise noted (when we investigate their impact on the system performance). 

\subsection{Micro-benchmark}
\label{ss:microBenchmark}

We start with performing micro-benchmarks to evaluate the effectiveness of each function module in \systemname.

\subsubsection{Chirp Synchronization}
\label{exp_sss:chirpSync}
Experiments in this section aim to i) evaluate the overall performance of the two-step chirp synchronization algorithm, and ii) understand the relationship between synchronization delay and number of slave radios.

\paragraph{i). The accuracy of chirp synchronization algorithm}. We synchronize chirp signals from all 24 slave radios using the two-step chirp synchronization algorithm.
We repeat this experiment 100 times and plot the CDF of the residual time offsets before and after applying our algorithm in Figure~\ref{fig:snc1}.  Without chirp synchronization, the median and maximum time offsets are 3630 and 8182 samples, respectively.
These two values drop to around 86 and 168 samples after the first-step chirp synchronization, and 0.4 and 0.9 sample after the second-step chirp synchronization. The trend clearly demonstrates that our chirp synchronization algorithm can effectively calibrate out the initial time offset among radios.

\paragraph{ii). Synchronization delay vs. slave count}. We then evaluate the chirp synchronization delay (termed as delay) under a different number of slave radios.
We repeat this experiment 100 times in each setting and plot the results in Figure~\ref{fig:snc2}.
We observe the delay increases with the number of slave radios.
Specifically, the delay is below 1.4~s when we have six slave radios, 2.9~s for 12 slaves,  5.3~s for 18 slaves, and 6.4~s for 24 slaves.
Please note that our chirp synchronization needs to run only once, and the one-time delay of 6.4~s would not have a noticeable effect on the user experience.

\subsubsection{Cold Start}
\label{exp_sss:coldStart}

Experiments in this section aim to evaluate the cold start success rate and delay when we vary i) number of slave radios and ii) distance between the leader and the backscatter (termed as L--B distance).

\paragraph{i). Success rate and delay vs. slave count}. In these experiments, we insert a backscatter radio into a 10~cm-thick pork belly and cold start it using a different number of slave radios.
The leader radio is half a meter away from this pork belly.
We perform this experiment 100 times in each setting and plot the success rate and delay in Figure~\ref{fig:cs1}.
We observe the success rate is low when we use only four slave radios.
This is as expected since beamforming power of four slave radios is too low.
The success rate soon jumps to 90\% as we double the number of slave radios.
It then approaches to 100\% when the slave count is larger than 15.
These results demonstrate the effectiveness of our cold start algorithm.
On the other hand, we see the delay of cold start decreases as we increase the number of slave participates.
The longest delay, however, is only 0.56~s.
These results clearly demonstrate the effectiveness of our cold start algorithm.

\paragraph{ii). Success rate  and delay vs. L--B distance}. We then test success rate and delay of our cold start algorithm in different L--B distance settings. In these experiments, we insert a backscatter radio into a 10~cm-thick pork belly and cold start it using 10 slave radios.
Results are shown in Figure~\ref{fig:cs2}.
The success rate is around 100\% when the leader radio is close to the backscatter radio, \eg, 0.3~m and 0.5~m away.
It then decreases slightly to 96.7\% and further to 86.6\% as we increase the L--B distance to 1~m and 2~m, respectively.
We observe a significant performance degradation (from 96.7\% to 26.3\%) when the leader is 3~m away from the backscatter radio.
This is because the searching space is too large, hence the power of side lobes is not strong enough to wake up the backscatter radio.
As for cold start delay, we observe that it grows smoothly as we increase the L--B distance from 0.3~m to 2~m.
It then jumps to around 0.3~s and further to 0.35~s as we place the leader radio 3~m and 4~m away from the backscatter radio, respectively.
By default, the leader radio is placed less than 1~m away from the human body. This is very practical because in real life, the leader radio can simply be the smartphone carried by the target or the smartwatch on the target's wrist. These devices are usually less than 1~m away from the implant inside the body. 

\begin{figure*}[t]
\begin{minipage}[t]{0.24\textwidth}
     \centering
{\includegraphics[width=1.05\linewidth]{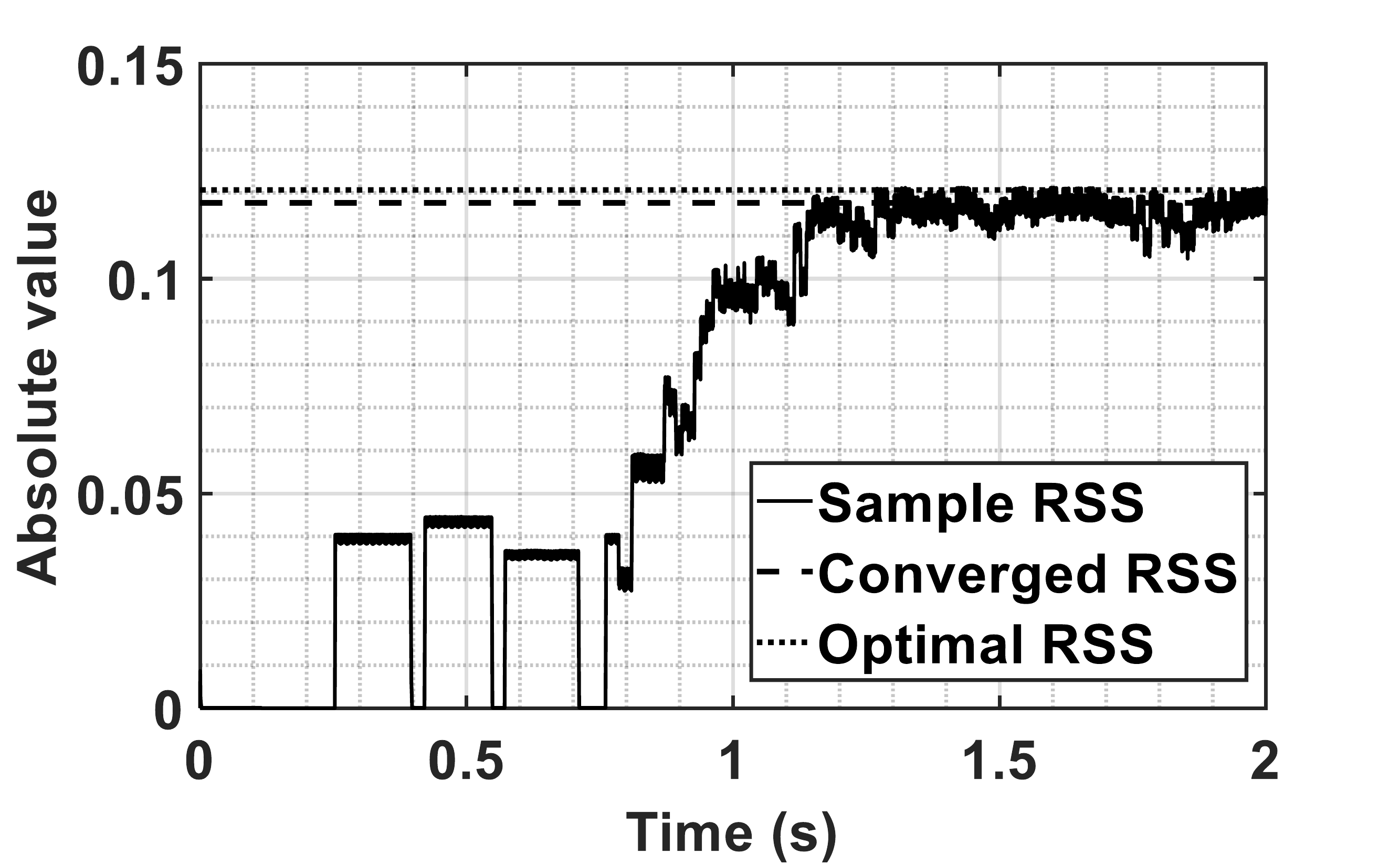}}
\caption{RSS during a beamforming episode.}
\label{fig:bf_vali}
    \end{minipage}\hfill
\begin{minipage}[t]{0.24\textwidth}
     \centering
{\includegraphics[width=1.05\linewidth]{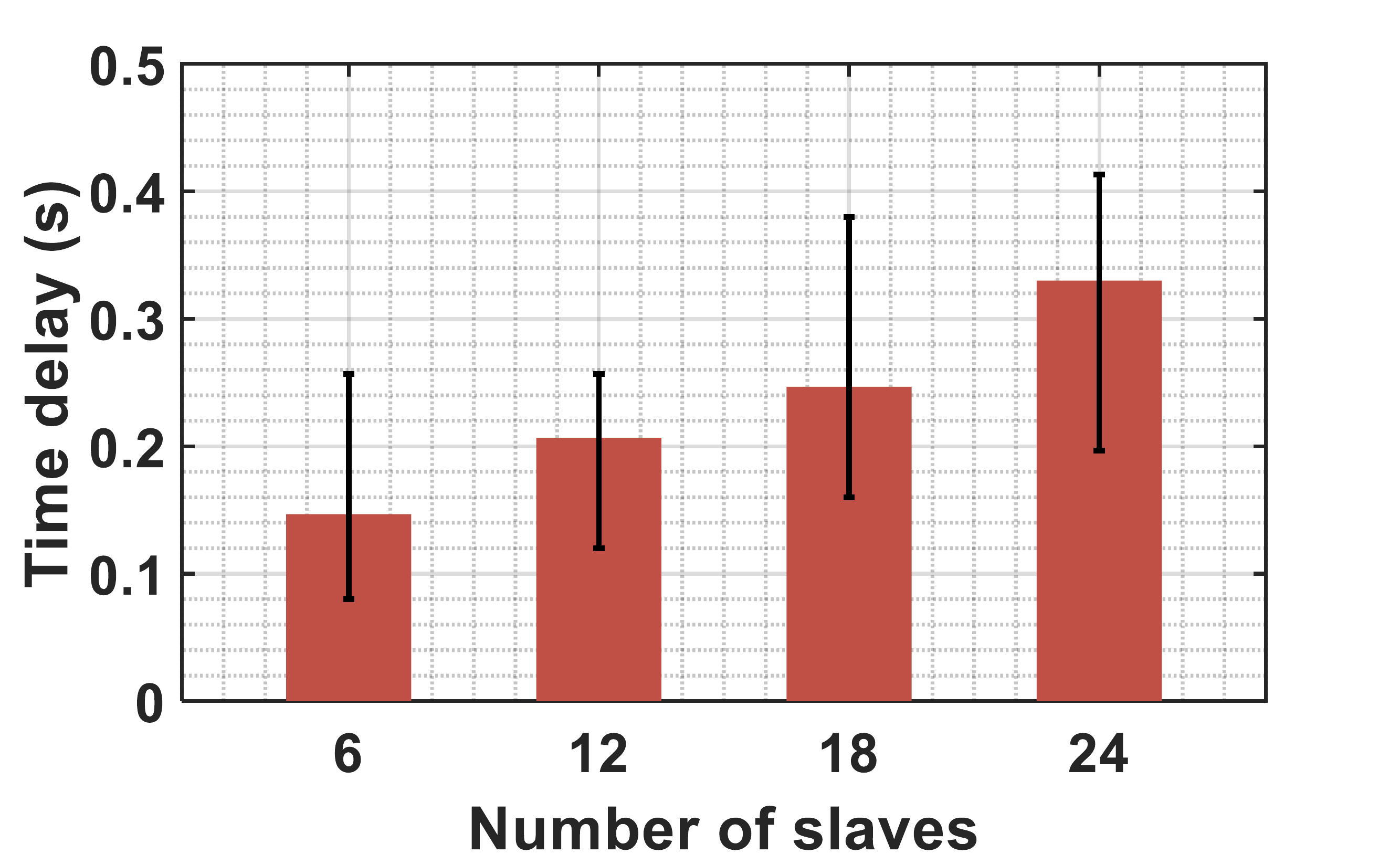}}
\caption{Beamforming delay vs. number of slave radios.}
\label{fig:bf2}
\end{minipage}\hfill
    \begin{minipage}[t]{0.24\textwidth}
     \centering
{\includegraphics[width=1.05\linewidth]{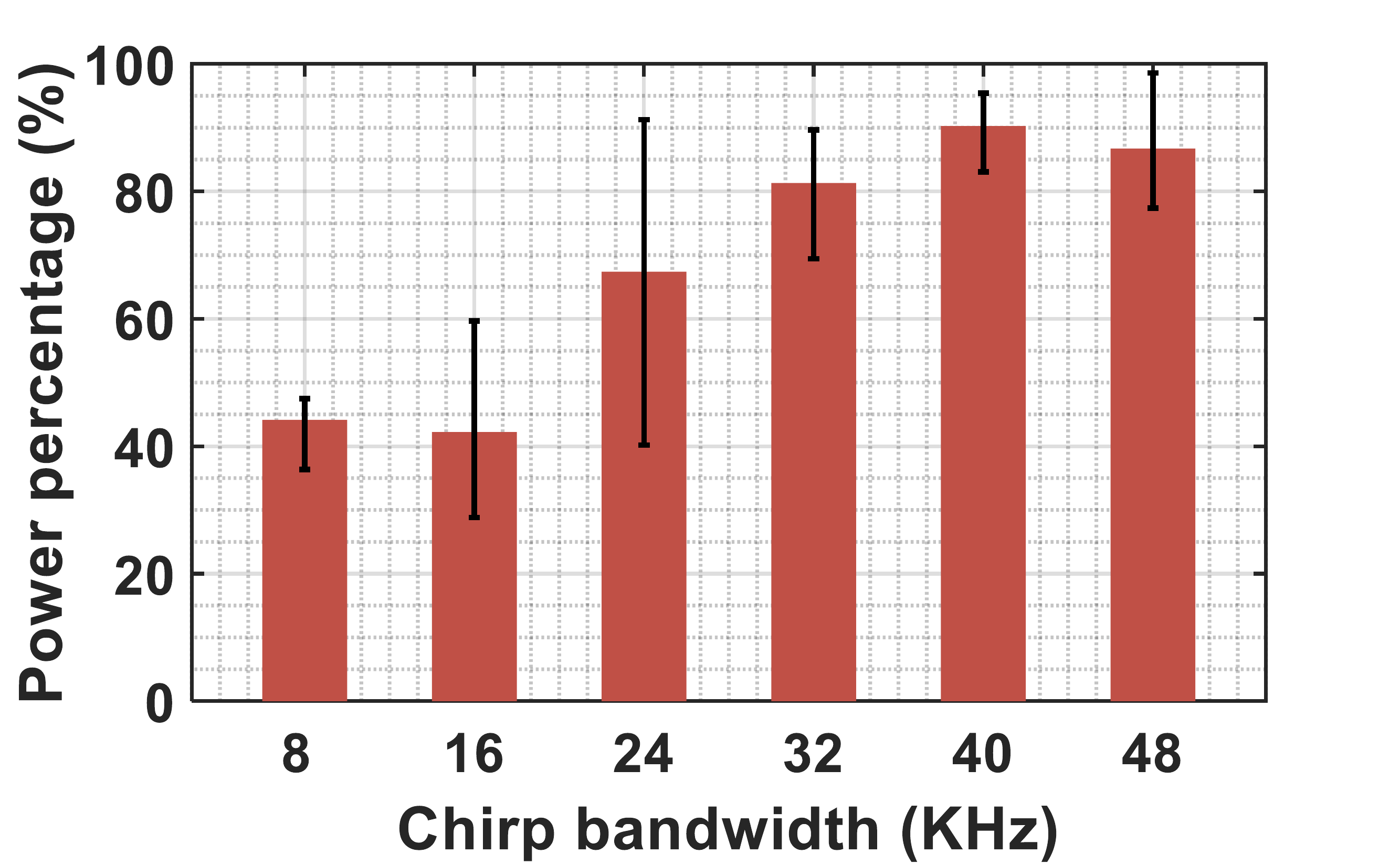}}
\caption{Power percentage vs. chirp bandwidth.}
\label{fig:bf_bw}
   \end{minipage}\hfill
     \begin{minipage}[t]{0.24\textwidth}
     \centering
{\includegraphics[width=1.05\columnwidth]{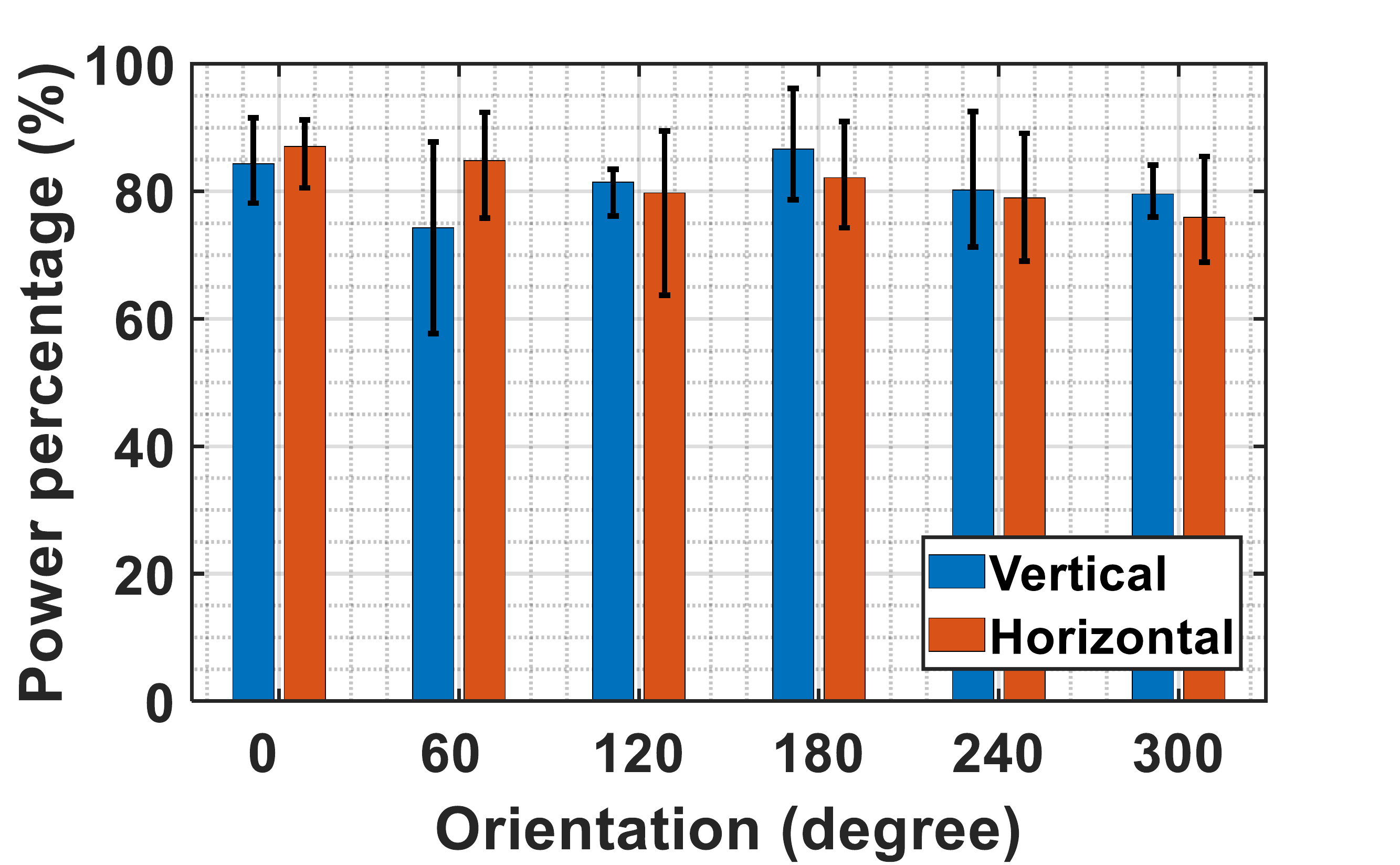}}
\caption{Power percentage vs. backscatter orientation.}
\label{fig:pp_r}
   \end{minipage}\hfill
\end{figure*}

\begin{figure*}
\begin{minipage}[t]{0.24\textwidth}
    \centering
    {\includegraphics[width=1.05\columnwidth]{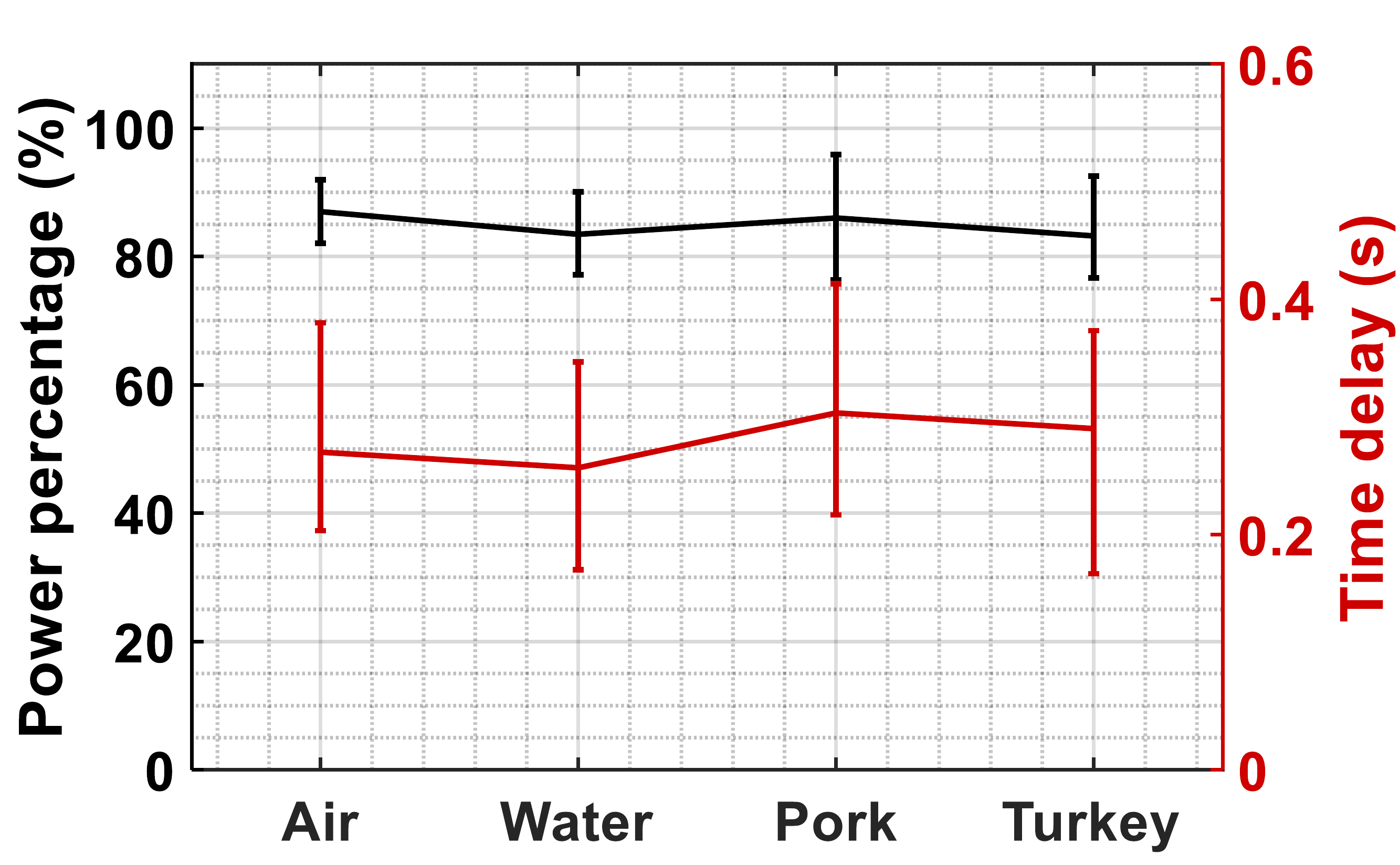}}
    \caption{Power percentage across different media.}
    \label{fig:media}
   \end{minipage}\hfill
\begin{minipage}[t]{0.24\textwidth}
\centering
{\includegraphics[width=1.08\columnwidth]{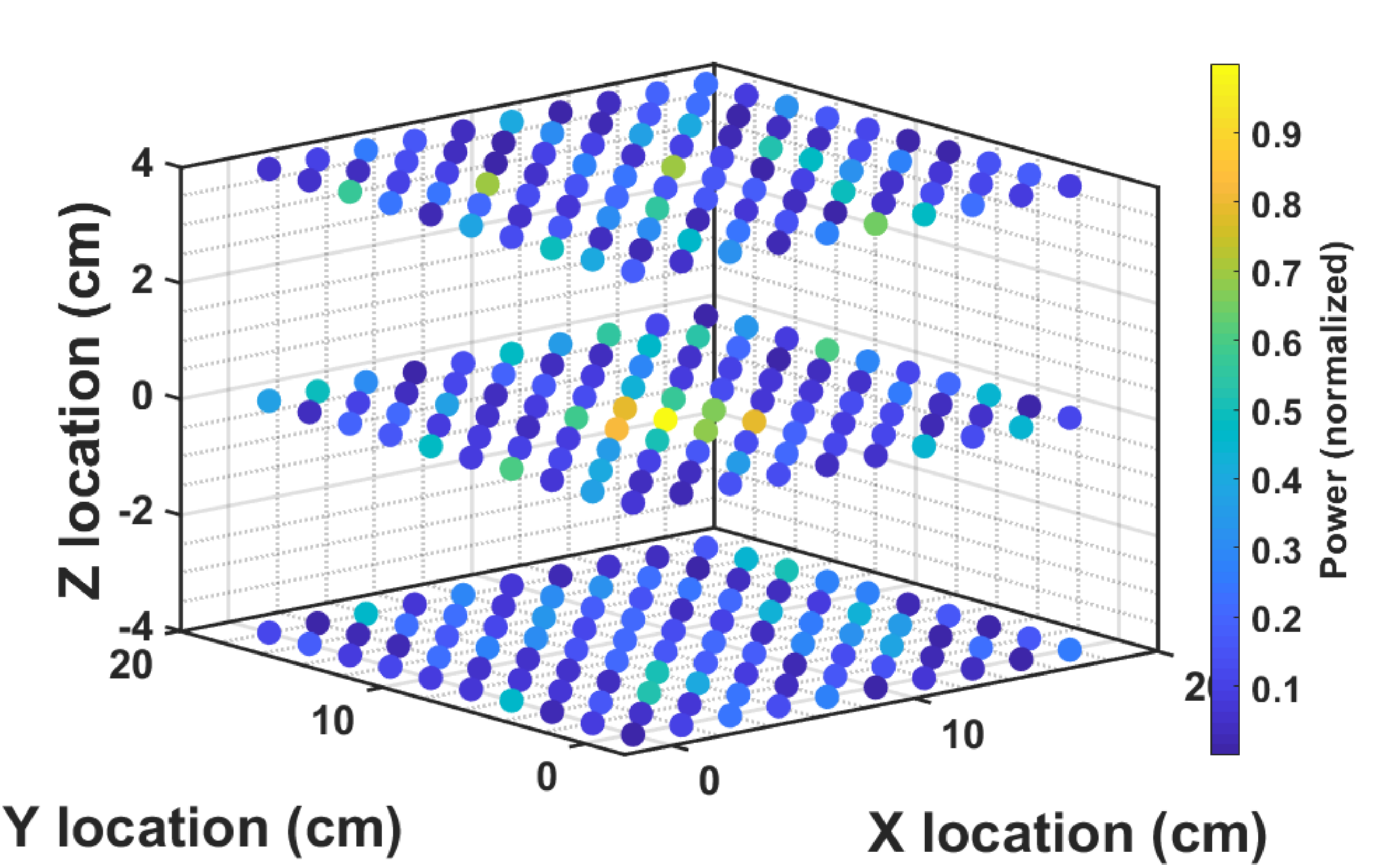}}
\caption{In-tissue 3D power distribution.}
\label{fig:energy_dist}
\end{minipage}\hfill
\begin{minipage}[t]{0.24\textwidth}
     \centering
    {\includegraphics[width=1.1\columnwidth]{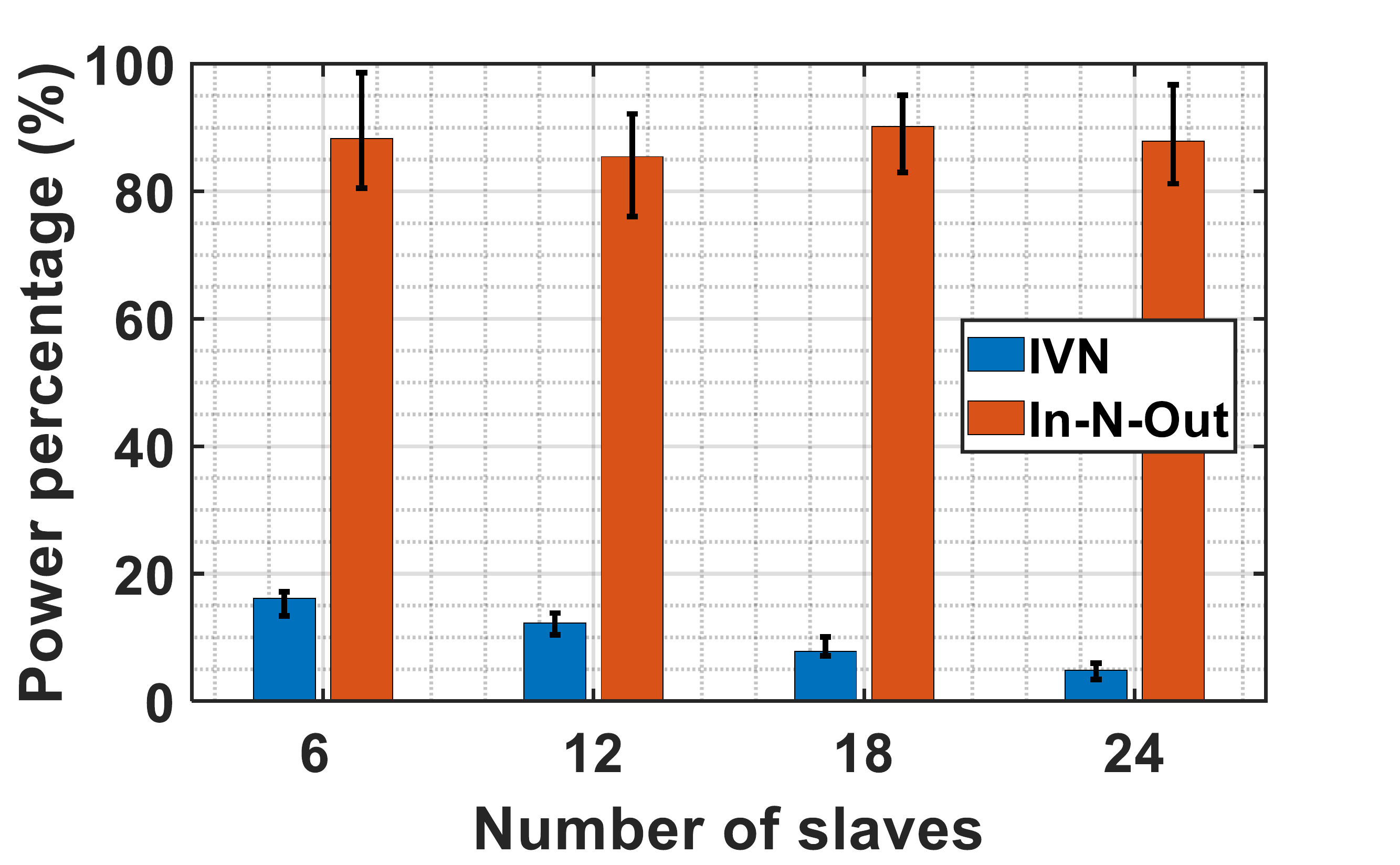}}
    \caption{Power percentage vs. slave count.}
    \label{fig:fs_bpp_1}
   \end{minipage}\hfill
\begin{minipage}[t]{0.24\textwidth}
     \centering
    {\includegraphics[width=1.1\columnwidth]{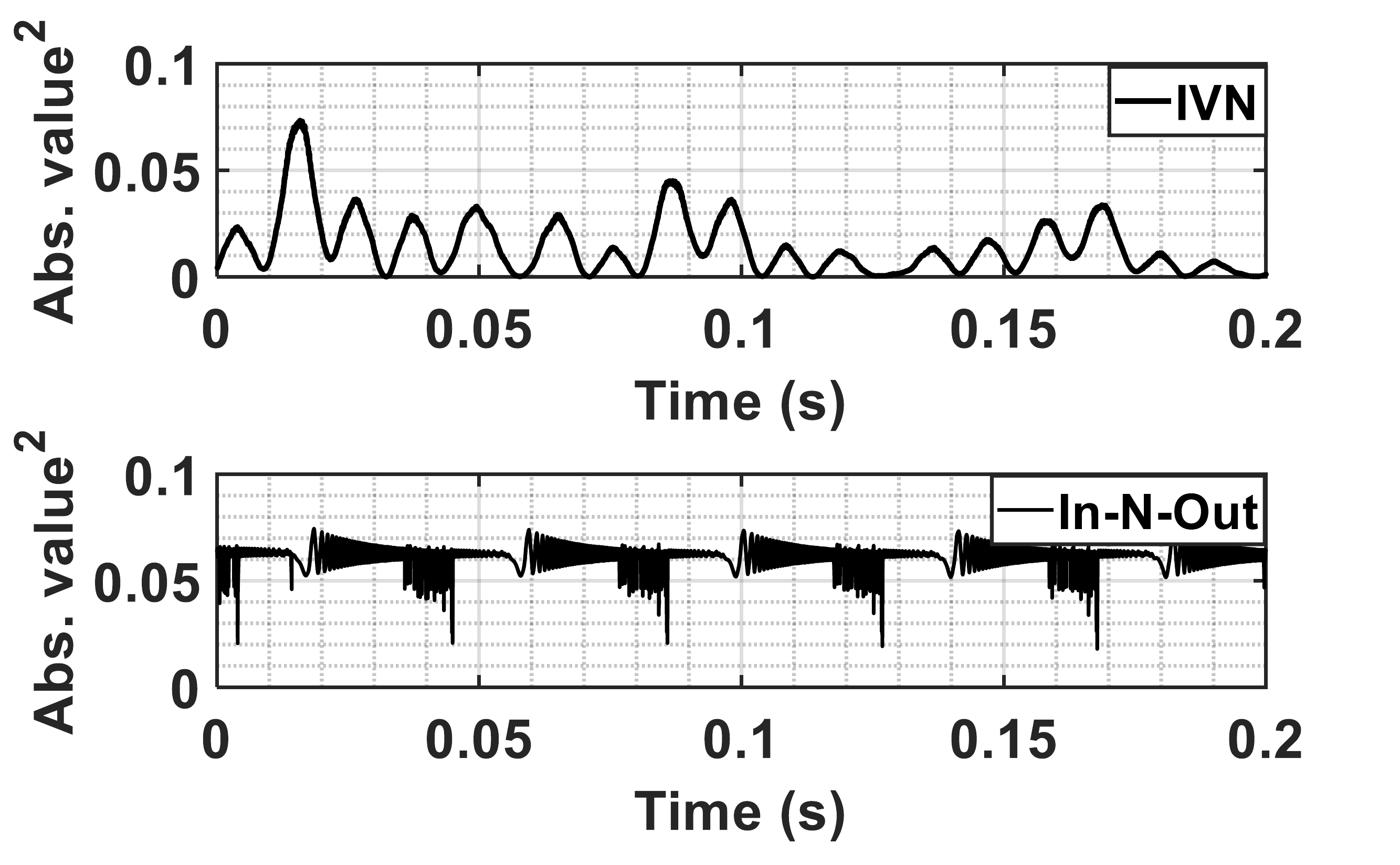}}
    \caption{Beamforming power samples.}
    \label{fig:fs_bpp_1_1}
\end{minipage}\hfill
\end{figure*}

\subsubsection{Beamforming}
\label{exp_sss:beamforming}

Experiments in this section aim to evaluate the delay and power gain of our beamforming algorithm in various parameter settings, \eg, different number of slave radios and different chirp bandwidths.
To measure the power gain gap between our beamforming algorithm and the optimal one,
we define a new metric, namely \emph{power percentage}, as the square of the ratio between the averaged beamforming amplitude (achieved by \systemname) and the optimal beamforming amplitude.
In reality, however, the optimal beamforming amplitude cannot be measured directly.
We thus start each slave radio at a time and record the received signal amplitude at the backscatter radio.
The summation of these signal amplitudes serves as an alternative to the optimal beamforming amplitude.
We also define the {\it beamforming delay} as the execution time of our beamforming algorithm until the beamforming power at the backscatter radio converges.

\paragraph{i). Close to optimal beamforming amplitude}. We measure the beamforming amplitude gap between our algorithm and the optimal one.
In this demonstrating experiment, we run our beamforming algorithm on three slave radios. 
The backscatter radio is inserted into a 10~cm thick pork belly placed 2~m away from each slave radio.
For a better illustration, we sequentially start these three slaves and measure their signal amplitude at the backscatter radio.
Figure~\ref{fig:bf_vali} shows the result.
We can see the beamforming amplitude grows rapidly and converges to a large value.
The convergence signal amplitude (dashed line) stays closely to the optimal beamforming amplitude (dotted line), with the average amplitude percentage of 96.5\%.
This result clearly demonstrates the high efficiency of the proposed beamforming algorithm.  

\paragraph{ii). Convergence delay vs. slave count}. 
Short convergence delay is crucial to our system, especially in mobile scenarios. 
We next examine the impact of slave count on the resulting beamforming delay.
The experiment setup follows the previous experiment.
We run the beamforming algorithm 100 times in each setting and plot the delay in Figure~\ref{fig:bf2}.
As we can see, the beamforming delay grows slowly as we increase the number of slave radios.
Specifically, the average delay is 0.15~s, 0.21~s, 0.25~s and 0.33~s with 6, 12, 18 and 24 slave radios, respectively. 
Further, we observe that though the beamforming delay fluctuates from experiment to experiment, the maximum delay is less than 0.41~s.
Hence, we believe that our iterative beamforming algorithm is fast enough for most of the in-body charging scenarios. 

\paragraph{iii). Power percentage vs. chirp bandwidth}. 
We then examine the impact of chirp bandwidth on the beamforming power percentage.
Similar to previous experiments, the backscatter radio is inserted into a 15~cm thick pork belly  placed 2~m away from 20 slave radios that are randomly picked from our testbed. 
We run the experiment 100 times in each setting and plot the achieved power percentage values in Figure~\ref{fig:bf_bw}.
As shown, the power percentage grows as we first increase the chirp bandwidth -- a higher chirp bandwidth improves the accuracy of the power inference algorithm.
An accurate power inference result further improves the beamforming efficiency.
Meanwhile, we observe that the power percentage increase rate decreases with the chirp bandwidth, indicating that the  marginal utility of the frequency-domain processing gain decreases.
Considering both trends, we set 40~KHz as the default chirp bandwidth setting.

\paragraph{iv). Power percentage vs. backscatter radio orientation}. 
We further examine how the backscatter radio's orientation affects the beamforming efficiency, which also indicates our system's robustness against the radio placement. 
In these experiments, we rotate the backscatter radio (inside a 15~cm thick pork belly) horizontally and vertically from 0$^{\circ}$ to 300$^{\circ}$ and measure the power percentage at the backscatter radio.
We repeat the experiment 100 times in each rotation angle and plot the results in Figure~\ref{fig:pp_r}.
We observe that \systemname achieves a consistently high power percentage (an average of 83.5\%, minimum of 74.3\% and maximum of 87.1\%) in all rotation angle settings.
This is because the antennas in our system are placed in a distributed fashion and thus are insensitive to the backscatter radio orientation.

\paragraph{v). Sufficient power for commercial medical implants}. To examine whether the beamforming power achieved by \systemname is sufficient to charge commercial medical implant, we conduct a   survey on the power consumption of several representative medical implants~\cite{halperin2008pacemakers,amar2015power}, including pacemaker, cardiac defibrillator, neuro-stimulator, and controlled internal drug release (CIDR).
For comparison, we also calculate the average beamforming power (in $\mu$W-scale) achieved by \systemname.
Table~\ref{tab:comp} summarize the result.
We measured 107~$\mu$W--617~$\mu$W (average 372~$\mu$W) beamforming power across the 18$\times$18~$m^2$ testbed area. 
The available power is higher than the power consumption of  pacemakers and cardiac defibrillators, and only slightly lower than some neuro-stim\-ulator and CIDR devices. 
We envision by shortening the signal prorogation path, 
such as deploying the system in smaller areas such as bedrooms or offices, \systemname would achieve a substantially higher beamforming power.

\begin{table}[t]
  \centering
  \begin{adjustbox}{width=1\columnwidth}
    \begin{tabular} {lrrrrr} \toprule[2pt]
      \makecell{\bfseries Device} & 
      \makecell{Pacemaker} & 
      \makecell{Cardiac defibrillator} & 
      \makecell{neuro-stimulator}&
      \makecell{CIDR} &
      \makecell{\systemname}\\ 
      \midrule[1pt]
      {\bfseries Power ($\mu$W)} & 
      10--100& 
      25--250& 
      40--500& 100--800 & 372\\
      \bottomrule[2pt]
    \end{tabular}
  \end{adjustbox}
\caption{{\bfseries Power requirements of several commercial medical implants}. \systemname can achieve average 372~$\mu$W by using 24 slave radios, which is sufficient to power up most of commercial pacemakers and cardiac defibrillators, as well as many neuro-stimulators and CIDRs.}\label{tab:comp}
\vspace{-15pt}
\end{table}

\begin{figure*}[tb]
     \begin{minipage}[t]{0.24\textwidth}
     \centering
{\includegraphics[width=0.92\columnwidth]{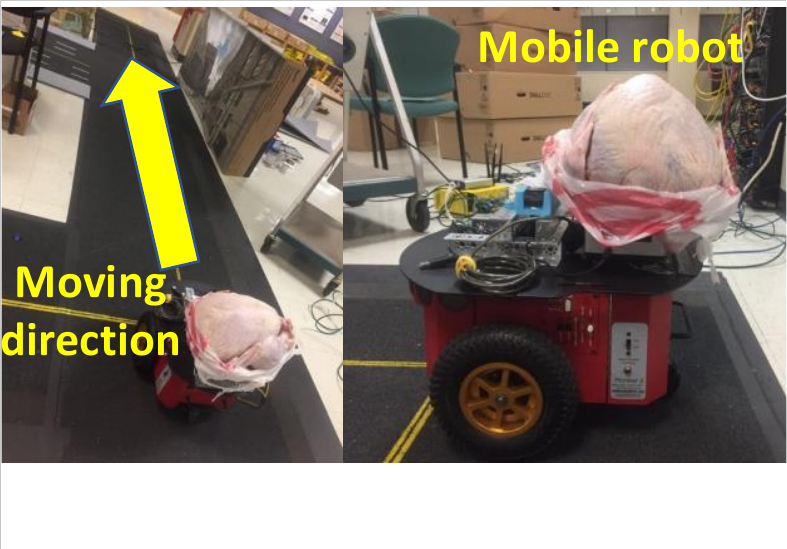}}
\caption{Experimental setup for mobile cases.}
\label{fig:mob1}
   \end{minipage}\hfill
   \begin{minipage}[t]{0.24\textwidth}
     \centering
{\includegraphics[width=1.1\linewidth]{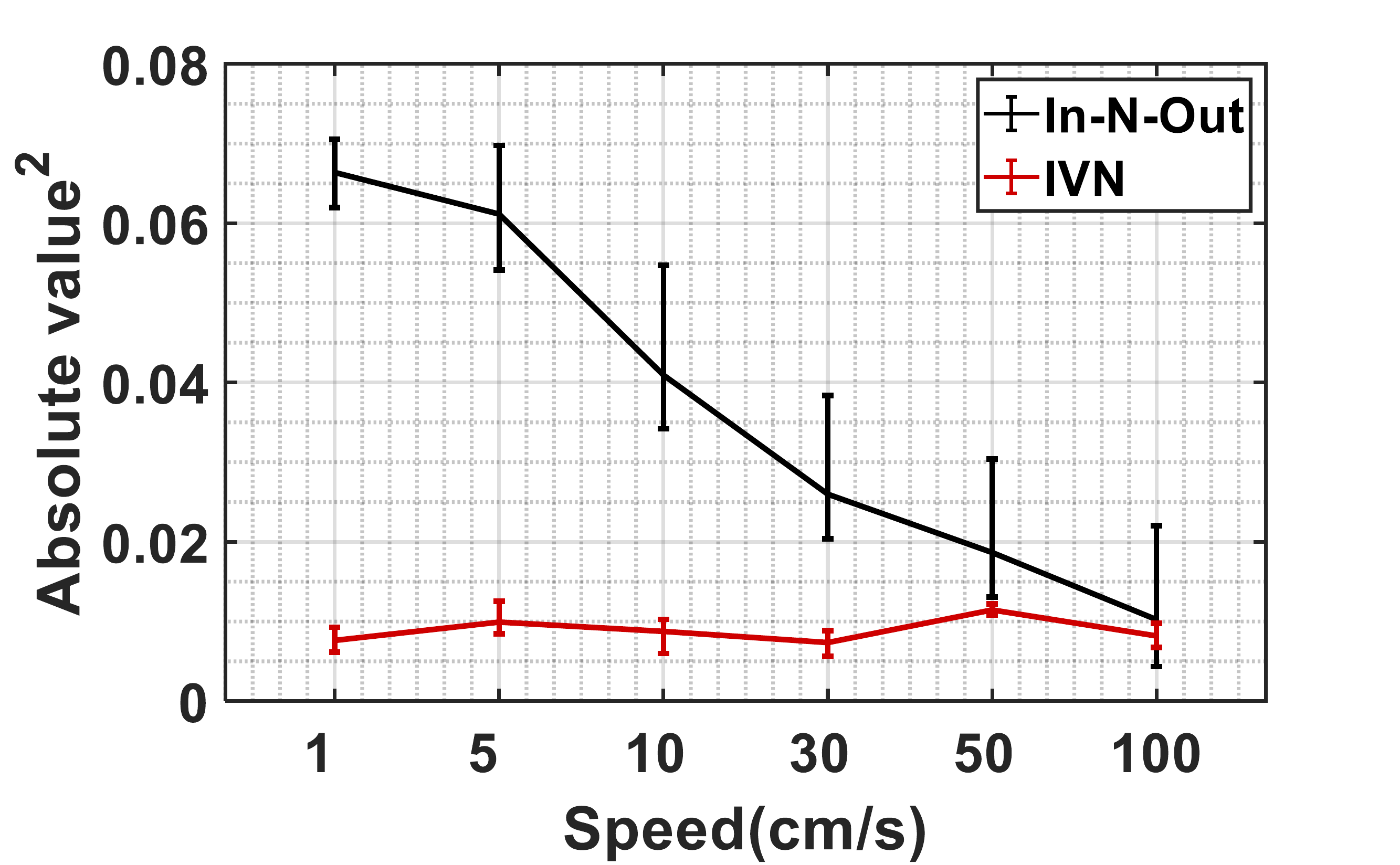}}
\caption{Average beamforming power vs speed.}
\label{fig:mob2}
   \end{minipage}\hfill
\begin{minipage}[t]{0.24\textwidth}
     \centering
{\includegraphics[width=1.1\linewidth]{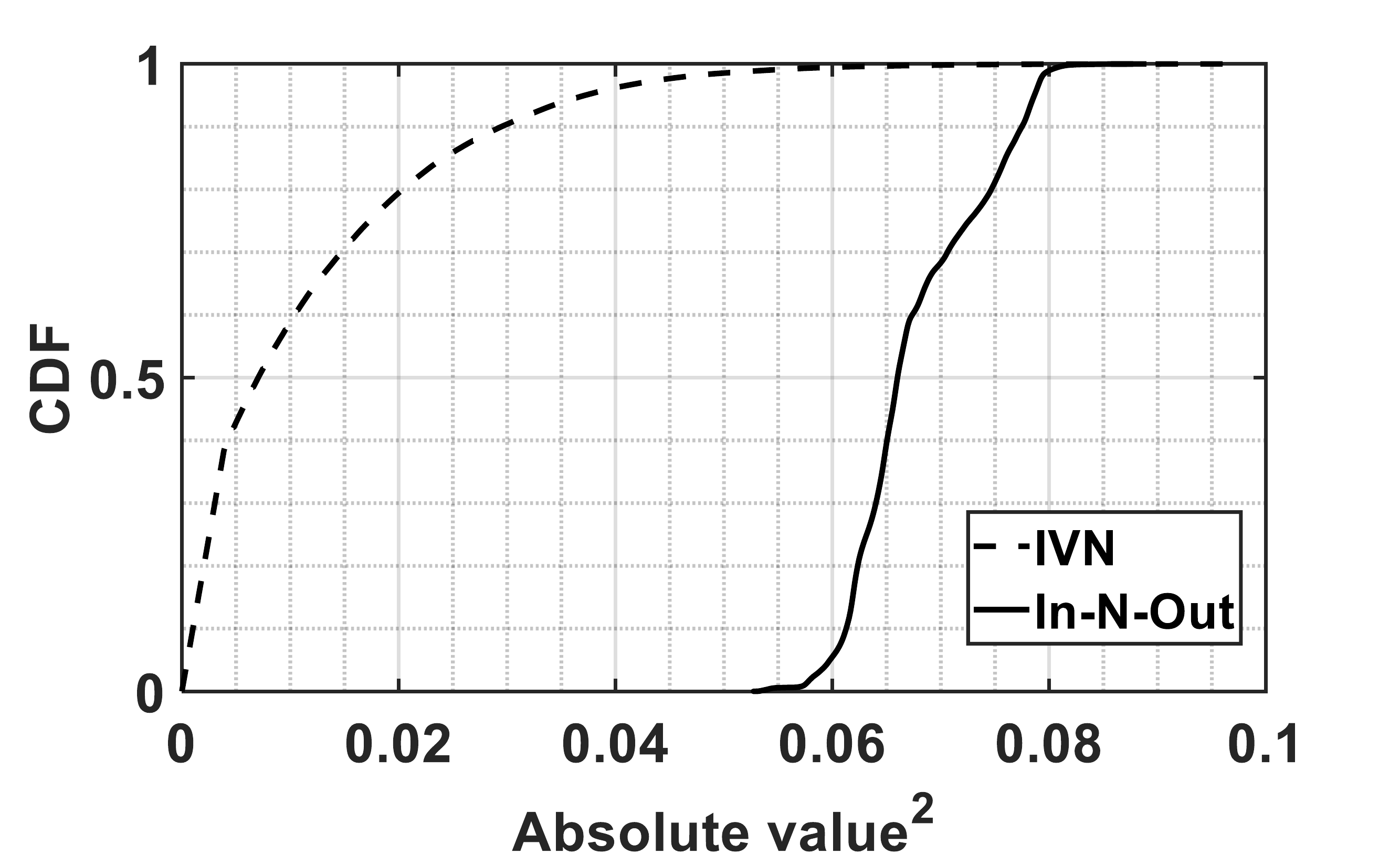}}
\caption{CDF of the beamforming power. ($v$ = 5~cm/s)}
\label{fig:mob3}
    \end{minipage}\hfill
   \begin{minipage}[t]{0.24\textwidth}
     \centering
{\includegraphics[width=1.1\columnwidth]{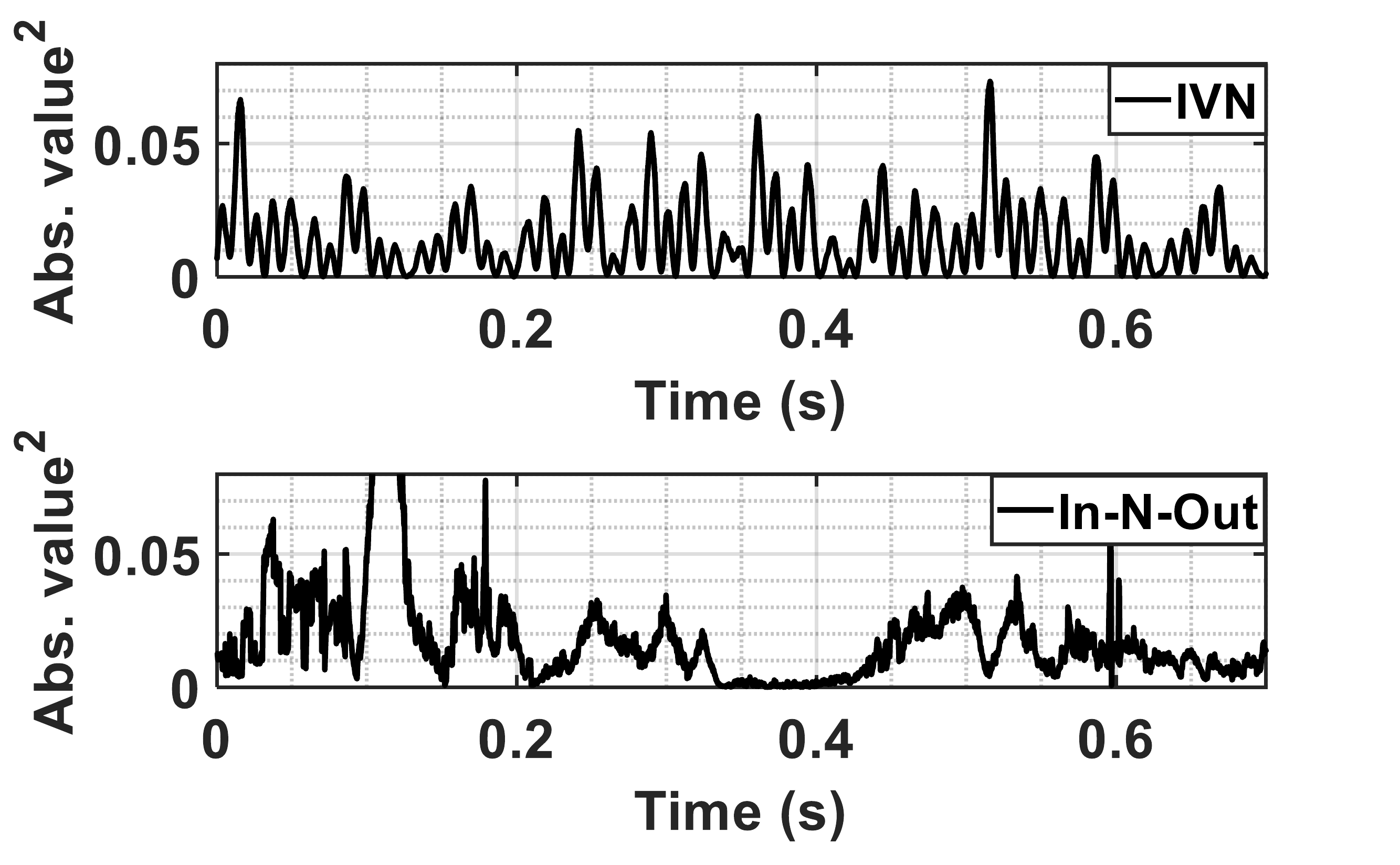}}
\caption{Beamforming power samples. ($v$=1~m/s)}
\label{fig:mob4}
   \end{minipage}\hfill
\end{figure*}

\subsection{Field Study}
\label{ss:exp_fieldStudy}

We next conduct field studies to evaluate the performance of \systemname in different mediums. Although there are a surge of inbody wireless charging competitors, we choose the state-of-the-art work IVN~\cite{yunfei} for comparison as IVN shares the most similar hardware setup with \systemname.
We carefully implement IVN and compare the performance of these two approaches in both stationary and mobile scenarios.

\paragraph{i). Impact of Medium}. 
We first examine whether our system can be used in other media.
In these experiments, we place the backscatter radio in four different media with significantly different channel characteristics, \ie, air, water, pork and turkey.
We measure the resulting power percentage and delay (excluding chirp synchronization) of our system in each setting.
24 slave radios are involved in these experiments.
As shown in Figure~\ref{fig:media}, \systemname achieves the highest average power percentage in the air (86\%), followed by 85\% in the water, 83\% in the pork belly, and finally 83\% in the turkey.
While the beamforming delays in these four media are slightly different, they are all below 0.41~s and would not cause noticeable delays in most of the cases.
These experiment results demonstrate that \systemname can be used to charge objects in various media.

\paragraph{ii). In-tissue Power Distribution}
We then examine \systemnameposs power distribution in deep tissues.
The backscatter radio is placed inside a 10~cm-thick pork belly.
Figure~\ref{fig:energy_dist} shows the power distribution measured across three slices (with 10~cm depth) of the pork belly.
The backscatter radio is placed at (8, 8, 0).
We observe a clear hot spot around the backscatter radio (with a radius around 2~cm) in the 3D space where the beamforming power is the highest.
The power at other locations, however, stays at a relatively low level.
The average power at the hot spot is 10.3$\times$ higher than the average power measured at the other locations.
This result clearly demonstrates that \systemname can successfully concentrate the beamforming energy to a tiny energy spot in a non-uniform medium like pork belly.

\paragraph{iii). Comparison with IVN in stationary cases}. 
We first compare the power gain of \systemname and IVN in stationary cases.
In these experiments, we insert a backscatter radio into a 10~cm thick pork belly and place them on a stationary table.
We then vary the number of slave radios from 6 to 24 and measure the power percentage achieved by both \systemname and IVN.
The experiment setup stays the same as the setup in the previous experiment.
We repeat this experiment 100 times in each setting and plot the results in Figure~\ref{fig:fs_bpp_1}.
As shown, \systemname achieves a consistently higher power percentage than IVN.
Specifically, when we have 6 slave radios, the average power percentage achieved by \systemname is 87.32\%, 5.4$\times$ higher than that achieved by IVN (16.2\%).
When we triple the number of slave radios (18), \systemname achieves 12.8$\times$ higher power percentage than IVN.
This gap further increases (18.1$\times$) as we use 24 slave radios.

To better understand the performance gap, we profile the instantaneous beamforming power of these two approaches and show the result in Figure~\ref{fig:fs_bpp_1_1}.
Both IVN and \systemname can achieve high beamforming power, but
IVN only achieves  high power levels at some time points. 
Its power level in most of the time, including the charging period, stays rather low, leading to a low average power level.
In contrast, the beamforming power achieved by \systemname is rather consistent, hence a much higher average power level. 

\paragraph{iv). Comparison with IVN in Mobile Cases}. \label{exp_sss:mobile}
We further conduct the performance comparison in mobile cases where the charging target moves around during the charging process.
In these experiments, we put a backscatter radio inside a 21~lb turkey.
The turkey is then fixed on a Pioneer-p3dx robot~\cite{alf} running ROS (Robot Operating System)~\cite{quigley2009ros}.
Figure~\ref{fig:mob1} shows the mobile experiment setup.
We use 10 slave radios and measure the received power level at the backscatter radio while the robot moves around.
The experiment is repeated 100 times in each speed setting.
Figure~\ref{fig:mob2} shows the average power achieved by \systemname and IVN.
When the robot moves at a relatively slow speed (\eg, 1~cm/s and 5~cm/s), \systemname outperforms IVN by 7.4$\times$ and 5.3$\times$, respectively.
To understand this difference, we plot the CDF of the beamforming power of these two systems when the robot moves at 5~cm/s.
The result is shown in Figure~\ref{fig:mob3}.
For \systemname, we find its power level stays rather consistent, with the lowest and highest power level of 0.053 and 0.089.
In contrast, the power level variation of IVN is much larger, with its 90\% percentile value below 0.029.
This result demonstrates that \systemname is agile enough to handle the target's slow movement. \eg, moving with the blood flowing.

As we increase the speed, the performance gap between these two systems decreases. Specifically,
when the robot moves at 1~m/s, the two approaches deliver similar power levels. 
To understand this trend, we randomly select a 0.8~s time window and measure the instantaneous beamforming power levels, as shown in Figure~\ref{fig:mob4}. The results show that at such a high speed, the power levels by both approaches vary drastically.
However, we expect that lower movement speeds such as 1~cm/s and 5~cm/s are much more commonplace than speeds as high as 1~m/s in medical implant charging scenarios. We believe \systemname can handle such common cases successfully.

\vspace{-2pt}

\section{Related Work}
\vspace{-2pt}
\label{s:related_works}
Our system is related to wireless charging and backscatter, while quantitatively differ from either one.

\subsection{Wireless Charging in Bioelectronics}
\label{wcib}

Wireless charging in bio-electronics can be broadly divided into three categories: near-filed inductive coupling, far-field electromagnetic radiation, and others.

\paragraph{Near-field inductive coupling} exploits magnetic field induction effect to deliver energy between two coils~\cite{kurs2007wireless}.
Research works in this domain focus on inductive power link optimization~\cite{ramrakhyani2011design,kurs2007wireless,jow2007design,zhang2019capttery}, source-load decoupling~\cite{ramrakhyani2013design}, and multi-coil linkage design~\cite{jadidian2014magnetic,kiani2011design}.
While near-field met\-hod achieves satisfying power delivery efficiency, it requires the user to wear bulky coils and align them with the implanted coil~\cite{standStill}.
As a result, the users need to sit still for hours to have their implants fully charged.
Moreover, the charging efficiency of near-field methods drops significantly with the reduction of coil size, which limits their working range to less than a centimeter~\cite{arbabian2016sound,hannan2014energy}.
Hence the focus in this field has shifted towards
overcoming the coil misalignment problem and improving the system robustness.

\vspace{-3pt}
\paragraph{Far-field wireless charging} transfers power to the target  through electromagnetic radiation~\cite{falkenstein2012low, liu2014design, chow2010fully}, microwave radiation~\cite{karalis2008efficient}, or laser~\cite{iyer2018charging,iyer2019living}.
Compared to the near-field method, the far-field method 
supports wireless charging over a longer distance at the cost of lower wireless charging efficiency.
Research in this field focuses on RF diode and DC impedance optimization~\cite{falkenstein2012low}, antenna optimization~\cite{liu2014design}, and effective system implementation~\cite{chow2010fully}.
IVN~\cite{yunfei} introduces an opportunistic frequency-encoding method in hope of combining signals constructively at the medical implant.
However, IVN's beamforming power, for most of the time, is far below the maximum value it can potentially achieve. 
In contrast, \systemname aligns the phase of signals at the medical implant rapidly and keeps this coherent phase combining over the entire wireless charging period.
Hence it can continuously charge the medical implant with consistently near-optimal beamforming power. 
The different design principle of \systemname and IVN leads to a significant gap in power delivery efficiency: In-N-Out achieves 5.4 – 18.1$\times$ and 5.3 – 7.4$\times$ power gain over IVN in stationary and mobile case, respectively.

\systemname also builds upon past works that leverage one-bit phase alignment algorithm~\cite{fan2018energy,mohanti2018wifed} for wireless charging.  
Energy-ball~\cite{fan2018energy} adopts this algorithm to charge IoT devices where CSI is unavailable.
WiFED~\cite{mohanti2018wifed} employs this algorithm to realize near optimal power transferring and communication over Wi-Fi links.
However, both of these pioneer works assume the receiver has enough battery to assess the beamforming power and produce feedback signals, which is not true for the ultra-low power, energy-scarce medical implants.
Besides, the excessive link budget renders the feedback signal far below the noise floor, and thus fail the feedback signal detection and decoding on the transmitter side.
Accordingly, we cannot directly borrow these techniques for inbody wireless charging.

As another alternative, mid-field resonant power transfer that combines both near-field and far-field methods is proposed~\cite{poon2010optimal,ho2014wireless,kim2012wireless,kim2012wireless}.
While this method can work over longer distances in the free space, the working range in the human body is still constrained by the coil spacing. 
Although the focus of this review is on RF-based methods, there are also related works on leveraging ultrasound for power transfer~\cite{maleki2011ultrasonically,suzuki2002power,guida2016700,santagati2015u,santagati2017implantable,guida2019u}. In~\cite{guida2016700, suzuki2002power, maleki2011ultrasonically}, the authors demonstrate the feasibility and advantages of ultrasonic power charging for implanted devices in animal tissues and tissue mimicking materials. In~\cite{santagati2015u,santagati2017implantable,guida2019u}, the authors proposed end-to-end ultrasonic charging and communication systems, whereas they focus on protocol design, hardware form-factor minimization and system rechargeability.    
Although ultrasound-based methods achieve higher power transfer efficiency, they are still intrusive due to the requirement of placing the transmitter coil close to the receiver~(\eg, attach to the human skin).

\vspace{-2pt}

\subsection{Backscatter Communication}

Backscatter systems encode information on top of the remote carrier signal for ultra-low-power communication.
Recent studies on backscatter communication aim to improve the backscatter range~\cite{peng2018plora,talla2017lora,ma2017drone}, enhance the ubiquity~\cite{liu2013ambient,talla2015powering,kellogg2016passive,zhang2016enabling,hessar2019netscatter}, and enable new applications such as fine-grained localization~\cite{ma2017minding,shangguan2016design,vasisht2018body}, material identification~\cite{wang2017tagscan, ha2018learning}, and vehicle counting and localization~\cite{abari2015caraoke}.
\systemname takes advantage of the backscatter design to reduce the energy consumption of the medical implant. 

There are also several works studying wireless charging on backscatter node without explicit channel measurement~\cite{arnitz2013wireless,denicke2018backscatter}.
However, these works still require CSI measurement at transmitters, which is very challenging due to the severe signal attenuation in deep tissues.
\systemname addresses this challenge by precoding the carrier signal using chirp modulation and leveraging its unique processing gain in the frequency domain to improve the SNR of backscatter signals.
Additionally, \systemname introduces a new metric $P_{CCS(0)}$ to replace the unreliable power measurement, and uses this metric to guide the execution of beamforming algorithm.

\vspace{-2pt}

\section{Discussion}
\label{s:disc}

\systemname leaves room for further investigations, as discussed  below:

\paragraph{Reducing deployment cost} As a proof-of-concept, we implement \systemname on software-defined radios (\ie USRP) for fast-prototyping.
In the future we plan to customize the RF radio design to reduce the overall system cost. Given the light-weight computation tasks and narrow band communication nature of our system, one can customize the RF radio with a MSP430~\cite{430} MCU (\$2.09), a MAX2235~\cite{2235} power amplifier (\$2.16), a TI SN74LS624N~\cite{SN74LS624N} oscillator (\$3.94), two cc1100~\cite{cc1100} radio transceivers (\$3.65), and two W5017~\cite{W5017} antennas (\$7.25), which leads to a total cost around \$25. 


\paragraph{Scaling to multiple targets}. While the system design is illustrated in the single target settings, \systemname can be easily extended to multi-user scenario by introducing a MAC layer protocol such as time duplex multiple access (TDMA) or Frequency duplex multiple access (FDMA). We leave this for our future work.

\vspace{-0.1cm}
\section{Concluding Remarks}
\label{s:concl}

In this paper, we present the design, implementation, and evaluation of \systemname: a multi-antenna system that can continuously charge the medical implant at the near optimal beamforming power, even when the implant moves around inside the human body.
To achieve this, \systemname proposes a set of novel signal processing algorithms and a low-power, monotonic backscatter radio design.
We prototype \systemname on software defined radios and PCB boards.
The head-to-head comparison on a multi-antenna testbed demonstrates that \systemname achieves 5.4$\times$--18.1$\times$ and 5.3$\times$--7.4$\times$ average power gain over the state-of-the-art solution in stationary and low-speed mobile scenarios, respectively.
\systemname is the first step towards flexible wireless charging for medical implants. Moving forward, we will endeavor to address the following technical challenges: achieving optimal deployment of the antenna array, mitigating the impact of strong multi-path effects, charging multiple implants simultaneously, etc. We also plan to pursue subsequent clinical experiments for further validations.

\section*{Acknowledgements}
We thank the reviewers and our shepherd for their insightful comments. We also thank Dr. Lin Zhong for providing us useful feedback on the early version of this work. 
This work is supported by the Key Research Program of Frontier Sciences, CAS, Grant No.ZDBS-LY-JSC001 and partially supported by 2030 National Key AI Program of China Grant No. 2018AAA0100500.
Corresponding author: yanyongz@ustc.edu.cn

\let\oldbibliography\thebibliography
\renewcommand{\thebibliography}[1]{%
  \oldbibliography{#1}%
  \setlength{\itemsep}{1pt}%
}
\clearpage

\bibliographystyle{ACM-Reference-Format}
\bibliography{main}
\begin{appendix}
\section*{Appendix}

\section{$P_{CCS(0)}$ extraction}
\label{s:p_ccs0}
Let $x(t)$, $X(\omega)$, $p(t)$, $P(\omega)$, $n(t)$, and $N(\omega)$ be the reference chirp symbol, received backscatter signal, and channel noise in the time domain and frequency domain, respectively.
When the leader radio detects the backscatter signal, it multiplies incoming signals with the complex\hyp{}conjugate copy of the reference chirp: $(p(t)+n(t))x^{*}(t)$. Next we prove $P_{CCS(0)}$ is the peak value of $(p(t)+n(t))x^{*}(t)$, and it changes monotonically with the strength of backscatter input signal $p(t)$. 

Recall that multiplication in time domain is equivalent to  the convolution in the frequency domain, we can rewrite the former expression as:
\begin{equation}\label{eq:f_domain1}
   x^{*}(t)(p(t)+n(t)) = X^{*}(-\omega)\ast P(\omega)+X^{*}(-\omega)\ast N(\omega)
\end{equation}
On the other hand, the cross\hyp{}correlation can be represented as $X(\omega)\otimes P(\omega)=X(-\omega)\ast P(\omega)$~\cite{cross}. Hence we can rewrite above expression as:
\begin{equation}
	\begin{aligned}
	 x^{*}(t)(p(t)+n(t)) = X(\omega)\otimes P(\omega) + X(\omega)\otimes N(\omega)
    \end{aligned}
\end{equation}
where $X(\omega)\otimes N(\omega)$ is a constant noise term, $X(\omega)\otimes P(\omega)$ is the cross\hyp{}correlation between the reference chirp and the backscatter signal, and
$\omega$ is the cross\hyp{}correlation lag. 
When \systemname detects the incoming backscatter signal, it synchronizes the reference chirp with this backscatter signal by shifting the reference chirp at the frequency domain.
This operation leads to a cross\hyp{}correlation peak (if there is a peak) shows at the zero lag position. 
Without loss of generality, we neglect the noise term of the above expression and use $P_{CCS(\omega)}$ to represent the frequency domain cross-correlation between the incoming backscatter signal and the reference chirp.
%
\begin{equation}
	\begin{aligned}
	 P_{CCS(\omega)}&=X(\omega)\otimes P(\omega)\\
	 &=\frac{1}{2\pi}\int_{0}^{2\pi}(\sum\limits_{m=-\infty}^{\infty}x(-m)e^{i\sigma m})(\sum\limits_{k=-\infty}^{\infty}p(k)e^{-i(\omega-\sigma) k})d\sigma\\
	 &=\sum\limits_{m=-\infty}^{\infty}x(-m)\sum\limits_{k=-\infty}^{\infty}p(k)e^{-i\omega k}\frac{1}{2\pi}\int_{0}^{2\pi}(e^{-i\sigma(-m-k)})d\sigma\\
	 &=\sum\limits_{m=-\infty}^{\infty}x(-m)\sum\limits_{k=-\infty}^{\infty}p(k)e^{-i\omega k}.
    \end{aligned}
\end{equation}
Hence the $P_{CCS(0)}$ (zero lag peak strength) can be expressed as:
\begin{equation}
P_{CCS(0)}=\sum\limits_{m=-\infty}^{\infty}x(-m)p(m) 
\end{equation}
The above expression indicates that $P_{CCS(0)}$ is {\it linearly proportional} to the power of backscatter signal $p(\cdot)$.
The leader radio thus adopts the power change of $P_{CCS(0)}$ as the indicator of the power change received at the backscatter.

\section{Optimal phase searching bound estimation}\label{sec:OPS}

\begin{figure}[t]
\centering
{\includegraphics[width=.75\columnwidth]{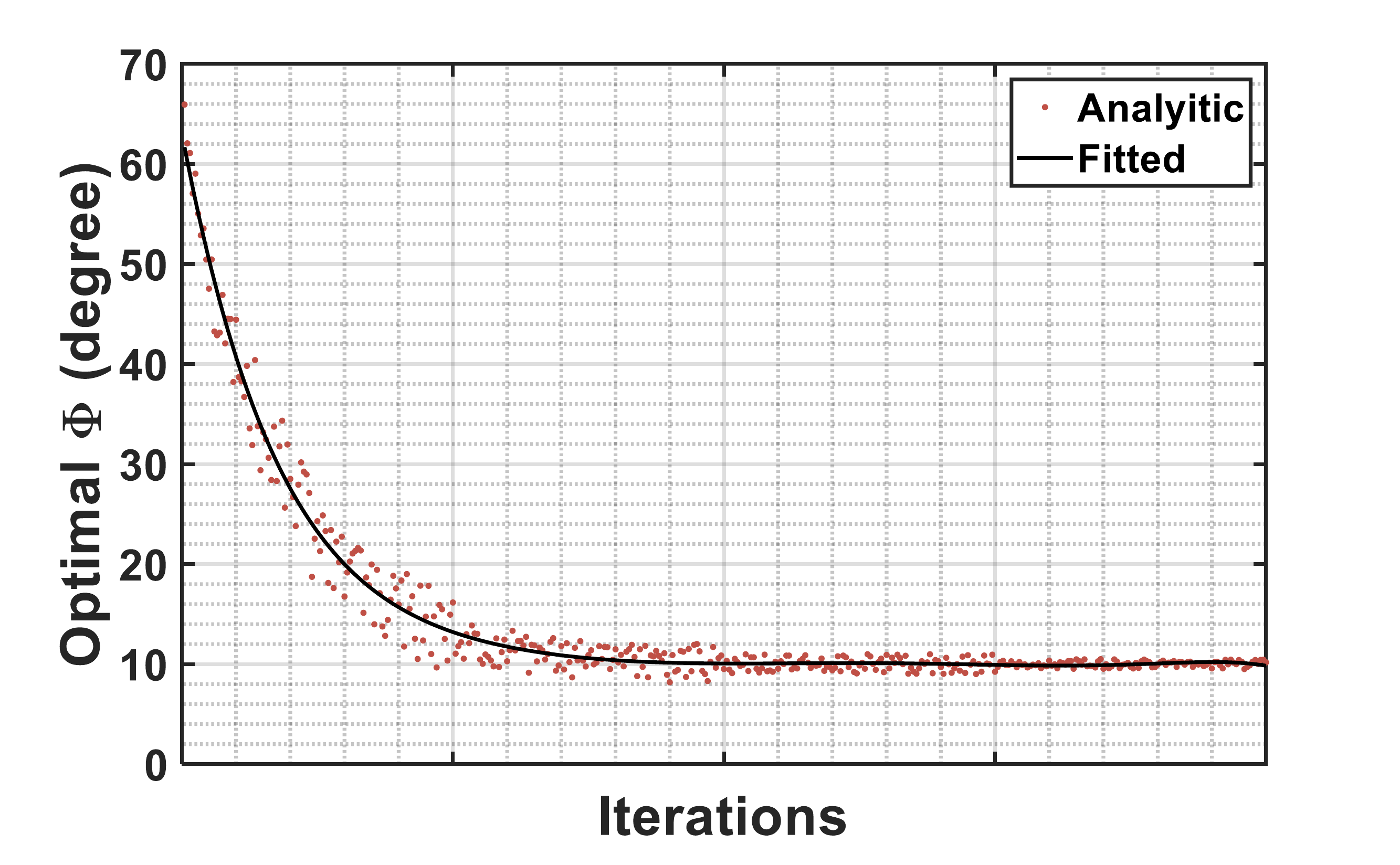}}
\caption{\textbf{Optimal phase searching bond and its corresponding 7 order polynomial fitting curve in the context of backscatter assisted beamforming.}}
\label{fig:result}
\end{figure}

According to the Proposition 3 in \cite{mudumbai2007energy}, the expected value of the beamforming amplitude after $n^{th}$ period is:   
\begin{equation}
	\begin{aligned}
	 &y[n+1]=y[n](1-p(1-C_{\Phi^{\circ}}))+\frac{\sigma_{1}}{\sqrt{2\pi}}e^{-\frac{(y[n](1-C_{\Phi}))^{2}}{2\sigma_{1}}}.\label{eqn:rss}
    \end{aligned}
\end{equation}
where \begin{equation}
	\begin{aligned}
	 &p=Q(\frac{y[n](1-C_{\Phi})}{\sigma_{1}}),\\
	 &\sigma_{1}^{2}=\frac{N}{2}((1-C^{2}_{\Phi})-\frac{I_{2}(\eta_{n})}{I_{0}(\eta_{n})}(C_{\Phi}^{2}-C_{2\Phi})),\\
	 &C_{\Phi}\doteq E_{\Phi}(cos \Phi_{i}),\\
	 &I_{k}(x)=\frac{1}{2\pi}\int_{-\pi}^{\pi}cos(k\phi)e^{xcos(\phi)}d\phi.
    \end{aligned}
\end{equation}
where $I_{k}(x)$ is the modified first-kind, n-order Bessel function, $\eta_{n}$ is characterized by $\frac{I_{1}(\eta_{n})}{I_{0}(\eta_{n})} = \frac{y[n]}{N}$.
$Q(\cdot)$ is the tail distribution function of the standard normal distribution.

In the context of the backscatter assisted beamforming system.
We first measure the backscatter signal power at different carrier signal power settings.
These results are then fitted using a nonlinear function, which is denoted by $P_{o}=\wp(P_{i})$.
Combining this nonlinear function with Equation~\ref{eqn:rss}, we have:
\begin{equation}
	\begin{aligned}
	 &y[n+1]=\wp(y[n](1-p(1-C_{\Phi}))+\frac{\sigma_{1}}{\sqrt{2\pi}}e^{-\frac{(y[n](1-C_{\Phi}))^{2}}{2\sigma_{1}}}).\label{eqn:rss1}
    \end{aligned}
\end{equation}
%

At each time slot $n$, we can calculate the optimal distribution of phase searching
bound $g_{n}(\Phi_{i})$ by solving the following optimization problem:
\begin{equation}
    \argmax\limits_{g_{n}(\Phi_{i})} (y[n+1]-y[n])    
\end{equation}
The problem of choosing an optimal distribution of phase searching bound is equivalent to the problem of finding the optimal variation of the phase searching bound. 
Given the fitted power function $P_{o}=\wp(P_{i})$, We compute the optimal phase searching bond at each iterations and plot the result in Figure~\ref{fig:result}.
To minimize the jitters, we then fit this analytic result using a high order nonlinear polynomial curve function $\Phi=P(n)$.
This function is then employed for setting the phase searching bound in each beamforming iteration.

\end{appendix}
\end{document}